\documentclass[11pt]{article}
\usepackage{cleveref}
\usepackage{color}
\usepackage{graphicx,amssymb,amsfonts,amsmath,amssymb,amscd,amstext}

\pdfoutput=1
\def\tr{\operatorname{tr}}
\def\calo{{\mathcal O}}

\crefname{section}{¤}{¤¤}
\Crefname{section}{¤}{¤¤}

\begin{document}

\begin{flushright}
YITP-SB-13-44
\end{flushright}

\begin{center}
\vspace{1cm} { \LARGE {\bf 
Losing Forward Momentum Holographically
}}
\vspace{1.1cm}

Koushik Balasubramanian and Christopher P. Herzog

\vspace{0.8cm}

{\it C.~N.~Yang Institute for Theoretical Physics \\ 
Department of Physics and Astronomy \\
Stony Brook University, Stony Brook, NY  11794}

\vspace{0.8cm}

\end{center}

\begin{abstract}
\noindent
We present a numerical scheme for solving Einstein's Equations in the presence of a negative cosmological constant
and an event horizon with planar topology.  Our scheme allows for the introduction of a particular metric source at the conformal boundary.  Such a spacetime has a dual holographic description in terms of a strongly interacting
quantum field theory at nonzero temperature.  
By introducing a sinusoidal static metric source that breaks translation invariance, we study momentum relaxation in the field theory.  In the long wavelength limit, our results are consistent with the fluid-gravity correspondence and relativistic hydrodynamics.  In the small amplitude limit, our results are consistent with the memory function prediction for the momentum relaxation rate.  
Our numerical scheme allows us to study momentum relaxation outside these two limits as well.

\end{abstract}

\pagebreak
\setcounter{page}{1}

\section{Introduction}

Through the AdS/CFT correspondence \cite{Maldacena:1997re,Gubser:1998bc,Witten:1998qj},
pure Einstein gravity with a negative cosmological constant describes a sector of
many strongly interacting, scale invariant quantum field theories (QFTs).  
The gravitational description allows one to compute properties of these QFTs at nonzero temperature and density,
perhaps providing some insight into strongly interacting real world systems in nuclear and condensed matter physics
(see \cite{Gubser:2009md,CasalderreySolana:2011us,Herzog:2009xv,Hartnoll:2009sz} for reviews).
Traditionally, the AdS/CFT approach has focused on systems at or near equilibrium.  Near equilibrium, we can make a further refinement into two cases: a) linear response where the departure from equilibrium itself is small and the system responds linearly and b) hydrodynamics where the departure happens over long times and distances such that any space-time point can be treated as if it were in local equilibrium.\footnote{%
Note that there are some examples where a hydrodynamic description seems to be valid even when the local thermal equilibrium assumption does not apply -- for instance, hydroization \cite{Chesler:2009cy,Chesler:2010bi}.}
Assuming a particle interpretation, long here means compared to scattering times and lengths.
Frontier questions are then those for which linear response and hydrodynamics both fail, for which
the departure from equilibrium is large and happens on short time scales and distances, 

The near equilibrium AdS/CFT approach has been quite valuable because strong interactions alone make it difficult to use traditional field theoretic techniques, but we would like to go further and address some of the many interesting questions surrounding far from equilibrium behavior.
What does turbulence look like when the driving happens on short time scales and wavelengths? How do shock waves behave in the limit when the shock becomes very thin?  How does fluid flow around a rough object when the scale of the roughness is of order or smaller than the mean free path?
More generally, when precisely do hydrodynamics and linear response cease to be valid and what should replace them?
AdS/CFT combined with numerical relativity provide methods to address these questions.

In this paper, we shall examine the case where the field theory is pushed away from equilibrium by a source for $g_{tt}$, the time-time component of the metric. More precisely, we will be solving the Einstein's equations that follow from the following four dimensional action:
\begin{eqnarray}
S = \frac{1}{2 \kappa^2} \int d^4x \sqrt{-g} \left( R - 2 \Lambda \right) \ ,
\end{eqnarray}
where the cosmological constant $\Lambda < 0$.  These equations admit a large family of solutions which have an asymptotic region that approaches four dimensional anti-de Sitter space.  This $AdS_4$ region in turn possesses a conformal boundary.  We will work in the Poincar\'e patch where this conformal boundary, in the absence of our $g_{tt}$ source, is conformal to three dimensional Minkowski space.
Through the AdS/CFT correspondence, the four dimensional spacetime is dual to a large class of three dimensional strongly interacting conformal field theories.  The metric $g_{tt}$ source in the field theory acts in gravity as a boundary condition for the metric at the conformal boundary.
When examining
linearized perturbations around a translation invariant equilibrium state, the system has more symmetry, and Einstein's equations reduce to ordinary differential equations.  In our far from equilibrium context, there will be no such simplification.

We plan for this paper to be the first of a series.  Here we detail our methods and provide a simple application.  In the future,
we will look at some more involved applications.
In the limit in which the field theory has a hydrodynamic description, the metric source allows us to recreate many classic fluid mechanics experiments.  By introducing a time dependent, random $g_{tt}$, we can drive a turbulent state.
We would like to know if such a driven state satisfies the Kolmogorov scaling law.  Some partial evidence for this scaling
was found in the decaying, non-driven turbulence of \cite{Adams:2013vsa}.\footnote{%
 See also refs.\ \cite{Carrasco:2012nf, Green:2013} for related work on turbulence in 2+1 dimensional relativistic flows.
}
While we do not know how to introduce boundaries with no-slip conditions in gravity, $g_{tt}$ allows us to do the next best thing.  For example, we can make a numerical wind tunnel where we construct a region with a different value of $g_{tt}$ and then drag it through the fluid at a velocity of our choosing.
For supersonic flows,
such an experiment would be a worthwhile generalization of \cite{Gubser:2007ga,Chesler:2007an} where the source was effectively point-like and gravity was linearized.
It would also be interesting to find the dual gravity description of Stoke's drag of an object for a flow with low Reynold's number.

The application we focus on in this paper is very similar in spirit to the Stoke's drag problem.
Starting with uniform subsonic fluid flow,
we measure the momentum loss in the presence of a metric source
$g_{tt} = -1 - \delta \cos(kx)$.  By making $g_{tt}$ depend only on $x$, our gravity simulation becomes effectively 2+1 instead of 3+1 dimensional, allowing us to get results more quickly.\footnote{%
 The related problem of a statonary flow over a metric ``bump'' was considered in \cite{Figueras:2012rb}.
}

In the hydrodynamic regime ($k / T \ll 1$), it is possible to obtain a heuristic estimate of the momentum relaxation time scale using an analogy with Stoke's formula for drag. Note that an object moving in a viscous fluid in the absence of external body forces will come to rest due to the effect of viscous drag. When viewed from the rest frame of the object, the flow relaxes to zero velocity after a certain time. Using Stoke's formula for the drag coefficient $C_d$, we write
$$ C_d \sim {1\over Re} \sim {\eta\over \rho U L}$$
where $Re$ is the Reynold's number,  $U$ is the characteristic flow velocity, $L$ is the characteristic length scale of the object, $\eta$ is the viscosity and $\rho$ is the mass density of the fluid. Note that the drag coefficient for a sphere is $24/Re$.
Now to estimate the lattice induced momentum relaxation time scale $\tau$, we choose the object to be a lattice with spacing $L= 2\pi/k$. Note that the flow decelerates to zero velocity in a time scale $\tau$ due to presence of the obstacle; hence the characteristic velocity is $U = L/\tau$. Assembling the estimates yields
$$ {1\over \tau} \sim {1\over C_d} {\eta k^2\over \rho} \ .$$
In Stoke's formula, the drag coefficient $C_d$ is a dimensionless quantity which depends on the shape or aspect ratio of the object. In the case of a lattice, we expect $C_d$ to be a function of the lattice ``strength" $\delta$.

In this paper, we obtain the relaxation time scale for flows using relativistic hydrodynamic simulations and numerical general relativity. We show that for small $k/T$, both hydrodynamic and gravity results agree with the above heuristic argument.
When $\delta$ is small, the system is near equilibrium where we check our results using linear response (more specifically the memory function method \cite{Forster,Hartnoll:2008hs,Hartnoll:2012rj}).  These cross checks give us faith in the robustness and accuracy of our code as we move toward addressing more challenging problems.

Additionally,
our simple experiment may have some interesting condensed matter applications.  As mentioned above, much effort has gone into trying to apply AdS/CFT to certain strongly correlated condensed matter systems.  However,
simple AdS/CFT models typically possess translation symmetry which leads to zero DC resistivity $\rho_{DC}$, inconvenient for example if one wants to explain the linear rise with temperature in $\rho_{DC}$
of a high $T_c$ superconductor.
As a result, much effort has gone into investigating the effects of translation symmetry breaking in these models
(see for example \cite{Hartnoll:2008hs,Hartnoll:2007ih} for early papers).  Most of these papers break translation invariance using an external gauge field (i.e.\ a spatially dependent chemical potential) or a source for a scalar.  Our system is in some sense simpler and more economical as the translation breaking is provided by the metric itself, requiring the addition of fewer extra tuneable parameters.  Our code allows us to study the region where both $\delta$ and $k$ are large and standard techniques fail.
(More precisely, we mean $\delta \lesssim 1$ and $1 \lesssim k/T$.)
We also provide an analytical technique to compute the relaxation time scale for large $\delta$ and small $k$ using perturbative techniques in the hydrodynamic description. This perturbative result agrees with the numerics within numerical error. To our knowledge, the regime of large $\delta$ has not been explored before in the literature.
Though we focus on subsonic flows in the present paper, it is possible to use the current numerical framework to explore flows with large velocities as well.

The rest of the paper is summarized as follows. In section \ref{sec:numericalGR}, we will describe our numerical scheme for solving Einstein's equations in a spacetime that is asymptotically $AdS_4$ and that possesses an event horizon with planar topology.
In section \ref{sec:numericalhydro}, we describe our numerical scheme for solving the relativistic hydrodynamic equations.
Section \ref{sec:results} contains the results of our simple experiment.
In appendix \ref{sec:gorydetails}, we give more details how the numerical codes were implemented and cross checked.
Appendix \ref{sec:twolimits} describes how to determine the momentum relaxation rate from hydrodynamics 
in the small $\delta$ limit for arbitrary constant flow velocity.

\section{Numerical Gravity}
\label{sec:numericalGR}

In this section, we describe the method used for solving Einstein's equations in the presence of a negative cosmological constant. Our metric ansatz was inspired by ref.\ \cite{Winicour}, and our computational scheme is very similar to those described in refs.\ \cite{Chesler:2010bi,Winicour, Chesler:2013lia}.  (See also ref.~\cite{CaoHe}.)

We use a characteristic scheme \cite{Winicour} combined with spectral methods \cite{spectral,jpboyd,trefethen}.  For hyperbolic partial differential equations (PDEs), a characteristic evolution scheme reduces the PDEs to ordinary differential equations (ODEs) along the characteristics. In numerical relativity, the characteristic formalism  is based on a choice of frame of reference where outgoing null cones evolve from an initial null cone. 

\subsection*{\it 1. Bondi-Sachs null coordinate system}
Our characteristic formalism in relativity is based on a geometry that is described by the Bondi-Sachs metric:
\begin{eqnarray}
ds^2 = - \left( e^{2 \beta} V z - \frac{h_{AB} U^A U^B}{z^2} \right) dt^2 - \frac{2 e^{2 \beta}}{z^2} dt \, dz 
- \frac{2 h_{AB}  U^B}{z^2} dt \, dx^A + \frac{h_{AB}}{z^2} dx^A \, dx^B \ .
\end{eqnarray}
We would like for this metric to describe a four dimensional space-time.  Thus $A,B = 1,2$.  We will fix 
$\det h = e^{4 \chi(t,x,y,z)}$ by choosing
\begin{eqnarray}
h = 
e^{2\chi} \left(
\begin{array}{cc}
e^\alpha \cosh \theta & \sinh \theta \\
\sinh \theta & e^{-\alpha} \cosh \theta 
\end{array}
\right) \ .
\end{eqnarray}
This coordinate system (up to the redefinition $z = 1/r$) is described on p 32 of \cite{Winicour}.

Vacuum Einstein's equations in the presence of a negative cosmological constant $\Lambda < 0$ are
\begin{eqnarray}
G_{\mu\nu} \equiv R_{\mu \nu}  -   \Lambda  g_{\mu\nu} = 0 \ .
\label{EE}
\end{eqnarray}
With our metric ansatz, 
the system is under-determined.  By redefining the radial coordinate, we can eliminate either $\chi$ or $\beta$.  
Motivated by the form of the boosted black brane metrics, 
we will eventually partially fix the gauge by taking 
\begin{eqnarray}
\chi(t,x,y,z) = \frac{1}{4} \ln \left(1  + 2 z^3 \chi_3 (t,x,y) \right) \ .
\label{chitest}
\end{eqnarray}
The remaining gauge degrees of freedom in $\chi_3$ will be used to place an apparent horizon at $z=1$, allowing us to truncate the region $z>1$ and solve Einstein's equations on a constant radial domain $0<z<1$.  (We choose the radial coordinate such that $z=0$ corresponds to the conformal boundary.)\footnote{%
At this point our numerical strategy differs somewhat from \cite{Chesler:2010bi,Chesler:2013lia}.  Those authors chose to fix
$\beta$ instead.
}

\subsection*{\it 2. Apparent horizon}
\label{sec:apphor}
The fact that light cannot escape an apparent horizon classically allows us to restrict the computational domain to the region outside the apparent horizon.  Since the null rays on the apparent horizon are all converging, information behind the horizon is not propagated into the computational domain. 
We use a coordinate system where the apparent horizon is located at a fixed radial coordinate.
Spectral methods are much simpler to implement when the domain is a box of fixed size independent of time.

Before we proceed to the equations of motion, we would like to derive the constraints imposed by fixing the apparent horizon at $z=1$. 
Consider a vector field $\xi^\mu$ associated to the tangents of a congruence of null geodesics. In other words, consider a vector field that satisfies the conditions $\xi^\mu \xi_\mu = 0$ and  $\xi^\mu \xi_{\nu ; \mu} = 0$. 
Given the conditions on $\xi^\mu$, the expansion is given by $\Theta = {\xi^\mu}_{;\mu}$.  
An apparent horizon corresponds to a surface where $\Theta = 0$.  

We parametrize $\xi_\mu$ in terms of the gradient of a hypersurface $S=0$ orthogonal to the geodesic congruence and an arbitrary rescaling function $R$: 
$\xi_\mu = R S_{, \mu}$.  
Without knowing the precise form of $S$, we can solve for $S_{;t}$ using the null condition $\xi^\mu \xi_\mu = 0$.  
Note that the null condition also implies that $S^{;\mu} S_{;\mu} = 0$.  
The geodesic condition $\xi^\mu \xi_{\nu ; \mu} = 0$ implies the orthogonality $S^{;\mu} R_{;\mu} = 0$, allowing us to solve for $R_{;t}$.  

It turns out that we need one $\xi$ and one hypersurface equation $S$ for each 
instant of time $t$, so let's make them functions of $t$:
$\xi[t]$ and $S[t]$.  To find the apparent horizon at an instant of time $t_0$, we choose $S[t_0]$ such that $S[t_0](t_0,x,y,z) = z-1$.  Note that more generally $S[t_0](t,x,y,z) \neq z-1$.   Substituting $S[t]$ into $\Theta$, we conclude that
\begin{eqnarray}
\label{apphorcond}
\left.  z^2 (z \chi'- 1) V - D_A U^A- 2 \dot \chi \right|_{z=1} = 0 \ ,
\end{eqnarray}
where $\chi' = \partial_z \chi$ and $\dot \chi = \partial_t \chi$.

Note that a trapped surface need not always exist. Even if it exists, it may not be simply connected which is not very helpful for restricting the computational domain. In certain cases, the characteristic evolution could encounter caustics before they reach the apparent horizon. Some methods to deal with specific caustics (such as point caustics) in the characteristic formalism are known \cite{Stewart}. In the present paper, we work with smooth initial and boundary conditions that do not lead to formation of caustics within the domain. It is not clear if caustics can always be avoided in the characteristic formalism for asymptotically AdS solutions, but it seems unlikely. 

We will now present the equations of motion with the above choice of coordinates and gauge.
\subsection*{\it 3. Equations of motion}
\label{sec:eom}

Einstein's equations for the Bondi-Sachs metric have a convenient nested structure that allows for efficient numerical integration.
In the characteristic formalism, $\beta$, $U^A$ and $V$ on a time slice can be determined if the values of $\alpha$, $\theta$, and $\chi$ on that time slice are known. We will also need to assume knowledge of
$z=0$ boundary values of $\beta$ and $U^A$.
As we have not yet discussed a $z=0$ expansion of the solution, we save a detailed discussed of boundary conditions for later.

  The scheme first determines $\beta$ on a constant $t$ slice.
From the ${zz}$ component of Einstein's equations, we find
\begin{eqnarray}
\label{betaeq}
G_{zz}
&=&  \frac{4}{z}(-1 + z \chi') \beta' - 2 \chi''- \frac{1}{2} \left[4  (\chi')^2+ (\alpha')^2 \cosh^2 \theta + (\theta')^2 \right] \ ,
\end{eqnarray}
where $f' \equiv \partial_z f$.
This first order differential equation is solved for $\beta$ after specifying the boundary conditions on $\beta$ at $z=0$.

From $G_{Az}$,  we obtain
\begin{eqnarray}
G_{Az} &=& \frac{z^2 e^{-2 \chi}}{2} \pi_A'
- \frac{1}{z^2} e^{2 \chi} ( z^2 e^{-2 \chi} \partial_A \beta)'
+\frac{1}{2} h^{BC} D_C h_{AB}' - 2 D_A \chi' \ ,
  \label{eq:UA}
\end{eqnarray}
where we have defined
\begin{eqnarray}
\pi^A \equiv z^{-2} e^{2(\chi-  \beta)} h_{AB} (U^B)' \ .
\end{eqnarray}
This differential equation is solved in two steps, first for $\pi^A$ and then for $U^A$, and requires boundary data for $U^A$.

From the combination $h^{AB} G_{AB}$ we find
\begin{eqnarray}
\label{Veq}
h^{AB} G_{AB}&=& 
4 e^{-2(\beta+\chi)} z^2 \left( \frac{e^{2 \chi}}{z^2} d_t \chi \right)'
 - \frac{1}{2} e^{-4 \beta} h_{AB} (U^A)' (U^B)' - \frac{2 \Lambda}{z^2}
 \\
&& + {\mathcal R} -2 e^{-\beta -2 \chi} D_A e^{2 \chi} h^{AB} D_B e^\beta +
z^4 e^{-2 (\beta+\chi)} D_A e^{-2 \chi} \left( \frac{ e^{4 \chi} U^A}{z^4} \right)'  
 \ .
 \nonumber
 \end{eqnarray}
We have defined
\[
d_t \chi \equiv \dot \chi - \frac{z^2}{2} (z \chi' -1)V \ .
\]
 Here ${\mathcal R}$ is the Ricci scalar computed from the $2 \times 2$ metric $h_{AB}$ and $D_A$ is the associated covariant derivative.
 This equation is solved for $d_t \chi$ 
 by using the apparent horizon condition (\ref{apphorcond}) as a boundary condition at $z=1$.
 
 The remaining two linearly independent combinations of $R_{AB} -\Lambda g_{AB}$ allow one to solve for 
 \begin{eqnarray}
 d_t \alpha &=& \dot \alpha - \frac{z^3}{2} V \alpha' \ , \\
 d_t \theta &=& \dot \theta - \frac{z^3}{2} V \theta' \ .
 \end{eqnarray}
 To wit, we have
\begin{eqnarray}
  \label{eq:GAB}
\lefteqn{G_{AB} = e^{-2 \beta}
\Biggl( z e^{\chi} \left( \frac{e^{\chi}}{z} d_t \hat h_{AB}  \right)' 
- \frac{1}{2} e^{2\chi} h_{AB} 
 \tr [ h' \cdot (d_t \hat h)]
}
 \\
&&
-2 e^{\beta} D_A D_B e^\beta
-\frac{1}{2} e^{-2 \beta} h_{AC} h_{BD} (U^C)' (U^D)' 
+ \frac{1}{2} \left( \frac{h_{AB}}{z^2} \right)' D \cdot U
\nonumber \\
&&
- e^{2 \chi} (\hat h_{C(A})' (D^C U_{B)} - D_{B)} U^C) 
+h_{C(A} D_{B)} (U^C)' 
+2  \left( \chi' -\frac{1}{z} \right) D_{(A} U_{B)} 
\nonumber \\
&&
+h_{AB}' d_t \chi  + (D_C h_{AB}') U^C  
+ \left( 
 \frac{1}{2} e^{2 \beta} {\mathcal R} 
+2 z^2 e^{-\chi}  \left(\frac{e^\chi}{z^2} d_t \chi \right)' 
- \frac{ \Lambda}{z^2} e^{2 \beta} \right) h_{AB}
  \Biggr) \ ,
  \nonumber
\end{eqnarray}
where we have defined the normalized spatial metric $\hat h_{AB} \equiv e^{-2 \chi} h_{AB}$.
The two differential equations are solved with Dirichlet like boundary conditions at $z=0$.

The horizon value of $V$ can be obtained from the equation $G^z_t + U^A G^z_A$.  Using (\ref{apphorcond}) and (\ref{betaeq}), this equation reduces to the following elliptic equation on the apparent horizon $z=1$:
\begin{eqnarray}
\label{eq:Wapphor}
\lefteqn{G^z_t + U^A G^z_A|_{z=1} = - \frac{1}{2} D^2 V -  \frac{1}{2}\left[  e^{-2 \beta}  U' + 2 D \beta \right]  \cdot D V } \\
&&
- \left[ d_t \chi' + \frac{1}{2} D \cdot U - U \cdot D \chi' \right] e^{-2 \beta} V 
+e^{-2 (\beta-\chi)} (d_t \hat h_{AB} ) D^A U^B \nonumber \\
&&
+\frac{1}{4} e^{-2 (\beta-2 \chi)} \tr [(d_t  \hat h) \cdot  (d_t \hat  h)]
\nonumber \\
&&
+ \frac{1}{2} e^{-2 \beta} \left[ (D_A U^B) (D^A U_B) + (D_A U^B)(D_B U^A) - (D \cdot U)^2 \right] |_{z=1} \ . \nonumber
\end{eqnarray}
Note that this elliptic equation also plays a key role in the integration strategy described in \cite{Chesler:2013lia}.

Once the horizon value of $V$ is obtained, $\dot \chi$ at the horizon can be deduced from the apparent horizon condition (\ref{apphorcond}).  Having fixed the radial dependence of $\chi$ through a gauge choice, we can compute $\dot \chi$ everywhere.
The functional form of $V$ can then be reconstructed from the definition of $d_t \chi$.
With $V$ in hand, $\dot \alpha$ and $\dot \theta$ can be computed from the definitions of $d_t \alpha$ and $d_t \theta$.

Modulo boundary conditions which we will discuss momentarily, given $\dot \alpha$, $\dot \theta$, and $\dot \chi$, 
we can compute $\alpha$, $\theta$, and $\chi$  on the next time slice. 
This whole process can then in principle be iterated, carrying the solution forward an arbitrary number of time steps.

\subsection*{\it 4. Boundary expansion}

We make the gauge choice (\ref{chitest}) for $\chi$.
  Let us also assume that the sources for $\alpha$, $\theta$, and $U^A$ vanish.
Near the conformal boundary $z=0$, we find the expansions
\begin{eqnarray}
V &=& \frac{1}{z^3} ( V_0 e^{2 \beta_0} + V_2 z^2 + V_3 z^3 + O(z^4) ) \ , \\
\beta &=&\beta_0 + \beta_3 z^3 + O(z^6) \ , \\
U^A &=& U_1^A z + U_3^A z^3 + O(z^4) \ , \\
\alpha &=& \alpha_3 z^3 + O(z^4) \ , \\
\theta &=& \theta_3 z^3 + O(z^4) \ .
\end{eqnarray}
The function $\beta_0(t,x,y)$ is a source term.  The functions $V_3(t,x,y)$, $\alpha_3(t,x,y)$, $\theta_3(t,x,y)$, 
 and $U_3^A(t,x,y)$ are integration constants which determine the stress tensor in the dual field theory.
We find that
\begin{align}
V_2 &=  ( \partial_x^2 + \partial_y^2) e^{2 \beta_0} \ , \;\;\;
\beta_3 = - \frac{1}{2} \chi_3  
 \ , \; \; \;
 U_1^A = - \partial_A e^{2 \beta_0} \ .
 \end{align}
The $(t\mu)$ components of Einstein's equations each require $V_0 = -\Lambda/3$.  
In addition to the source term $\beta_0$, 
our boundary data consists of five parameters: $V_3$, $\alpha_3$, $\theta_3$, and $U_3^A$. 
Five parameters are exactly what is needed to describe a traceless stress energy tensor.
Defining the boundary stress tensor in the usual way \cite{Balasubramanian:1999re} as
\begin{eqnarray}
\label{Tabdef}
T_{ab} = \lim_{z \to 0} \frac{\sqrt{V_0}}{z}\left(K_{ab} - (K+2 \sqrt{V_0} ) \gamma_{ab}  + \frac{1}{\sqrt{V_0}} \left({\mathcal R}_{ab} - \frac{1}{2} {\mathcal R} \gamma_{ab} \right)\right) \ ,
\end{eqnarray}
where $\gamma_{ab}$ is the induced metric on a slice of constant $z$, $K_{ab}$ is the associated extrinsic trace, and ${\mathcal R_{ab}}$ is the three dimensional Ricci tensor.
We find that:
{
\begin{eqnarray}
T_{tt} &=& V_0  V_3 e^{2 \beta_0} - 2 e^{4 \beta_0} V_0^2 \chi_3\ , \\
T_{tA} &=& \frac{3}{2} V_0 U_3^A +  e^{-2 \beta_0} \partial_A [ e^{4 \beta_0} (\partial_C \partial^C \beta_0)  ] 
 \ , \\
T_{xx} &=& \frac{1}{2} e^{-2\beta_0} V_3 - \frac{3}{2}V_0 \alpha_3 + \frac{e^{-6 \beta_0}}{V_0}
\left[ \partial_x ((  \partial_x \dot \beta_0) e^{4 \beta_0})-\partial_y (( \partial_y \dot \beta_0) e^{4 \beta_0}) \right]  \\
&& - V_0 \chi_3 \ , \nonumber \\
T_{xy} &=&  - \frac{3}{2} V_0 \theta_3  + 
 \frac{e^{-6 \beta_0} }{V_0}
\left[ \partial_y ((\partial_x \dot \beta_0) e^{4 \beta_0})+\partial_x (( \partial_y \dot \beta_0) e^{4 \beta_0}) \right] 
 \ ,  \\
T_{yy} &=& 
\frac{1}{2} e^{-2 \beta_0} V_3 + \frac{3}{2} V_0 \alpha_3 - \frac{e^{-6 \beta_0} }{V_0}
\left[ \partial_x ((\partial_x \dot \beta_0) e^{4 \beta_0})-\partial_y (( \partial_y \dot \beta_0) e^{4 \beta_0}) \right] 
 \\
&& - V_0 \chi_3 
\ . \nonumber
\end{eqnarray}
}
Note that $\gamma^{ab} T_{ab} = 0$.

Given the derivation of (\ref{Tabdef}) from a variational principle, we are guaranteed that this stress tensor is covariantly conserved on the boundary $z=0$, i.e.\  $\nabla_a T^{ab}=0$.  (With $\beta_0 = 0$, the boundary is flat and we have the stronger condition $\partial_a T^{ab} = 0$.)  These conservation conditions impose the following differential relations on the five parameters:
\begin{eqnarray}
\label{consone}
 2 \partial_t (e^{-2 \beta_0} V_3) &=&   3 V_0 e^{-2 \beta_0} \partial_A (e^{2 \beta_0} U_3^A)
+ 2 e^{-2 \beta_0}
\partial_A \partial^A [e^{4 \beta_0} (\partial_C \partial^C \beta_0)]
  \nonumber \\
&&
- 4 \dot \chi_3
 \ , 
\\
\label{constwo}
3 \partial_t (e^{-2 \beta_0} U^1_3) &=&
e^{-4 \beta_0} \partial_x (V_3 e^{4 \beta_0}) 
- 3 V_0  \partial_x ( \alpha_3 e^{2 \beta_0})
- 3 V_0\partial_y ( \theta_3 e^{2 \beta_0})\\
&&
- 6 \chi_3 \partial_x e^{2 \beta_0} - 2 e^{2 \beta_0} \partial_x \chi_3 
\nonumber
 \ , \\
\label{consthree}
3 \partial_t (e^{-2 \beta_0} U^2_3) &=&
e^{-4 \beta_0} \partial_y(V_3 e^{4 \beta_0})
+ 3 V_0 \partial_y(\alpha_3 e^{2 \beta_0})
-3 V_0 \partial_x(\theta_3 e^{2 \beta_0}) \\
&&
- 6 \chi_3 \partial_y e^{2 \beta_0} - 2 e^{2 \beta_0} \partial_y \chi_3 
\nonumber
 \ .
\end{eqnarray}
These differential relations can also be obtained by systematically solving the equations of motion order by order in $z$. For example, 
considering the $(t\mu)$ along with the $(Az)$ components of Einstein's equations yields the three relations above in addition to
conditions on $V_4$ and $U_4^A$.

The six bulk equations of motion $G_{zz}$, $G_{Az}$, and $G_{AB}$ that we use are a subset of the ten Einstein's equations.
It is an interesting exercise to see how our integration scheme above guarantees that the four equations $G_{t\mu}$ are also satisfied.
With some effort, one can establish that $G_{tz}$ is a linear combination of $G_{zz}$, $G_{Az}$, $G_{AB}$ and their derivatives.
Although we do not use $G_{tt}$ and $G_{tA}$ in the bulk, we do use them to set boundary conditions.  In particular, 
we use them to propagate the boundary values of $U^A$ and also to derive the elliptic equation (\ref{eq:Wapphor}) used to set the value of $V$ at the horizon.
That $G_{tt}$ and $G_{tA}$ are satisfied everywhere then follows from a Bianchi identity, as we now argue.
Let us define
$$ H_{\mu\nu} = R_{\mu\nu} - {1\over 2 }R g_{\mu\nu}  - \Lambda g_{\mu \nu} \ . $$
The contracted Bianchi identity implies that 
 $$ \nabla_{\mu} H^{\mu}_{\nu} = 0 \implies {1\over \sqrt g} \partial_\mu \left(\sqrt{g} H^\mu_\nu \right) + \Gamma_{\nu \rho}^{\mu}H^{\rho}_\mu = 0 $$
 is satisfied as an algebraic identity.
 Now $H^z_z$ and $H^z_A$ depend linearly on $G_{tt}$ and $G_{tA}$ respectively.  Additionally, $H^t_z$, $H^t_A$, $H^B_z$ and $H^B_A$ are independent of $G_{tt}$ and $G_{tA}$.  
 Thus if $G_{tt}$ and $G_{tA}$ are satisfied at some point in $z$, integrating the Bianchi identity in the $z$ direction, 
 they must be satisfied everywhere in the interval $0\leq z \leq 1$.
The way we set the boundary conditions for $U^A$ guarantee that $G_{tA}=0$ are satisfied at $z=0$ so they must be satisfied everywhere.  The way we set boundary conditions for $V$ then guarantees that $G_{tt}$ is satisfied at $z=1$.  Thus $G_{tt}=0$ everywhere.  
A way of monitoring the accuracy of our integration scheme is to check how well $G_{tt}$ is satisfied at $z=0$, in other words
to monitor (\ref{consone}).  

\subsection*{\it 5. Marching orders}

We are now ready to specify precisely what fields we numerically integrate and which boundary conditions we apply.
We define new functions with subscript $s$.
\begin{eqnarray*}
\beta &=& \beta_0 - \frac{z^3}{2} \chi_3 + z^4 \beta_s \ , \\
U^A &=& - z \partial_A (e^{2 \beta_0}) + z^2 U_s^A \ ,  \; \; \;
\pi^A =-\frac{2}{z^2} \partial_A \beta_0 +  \pi^A_s \ , \\
V &=& \frac{1}{z^3} ( V_0 e^{2 \beta_0} +  z^2 V_s ) \ ,  \; \; \;
d_t \chi = \frac{e^{2 \beta_0}}{2 z} + 
\frac{z}{2} (\partial_x^2 + \partial_y^2) e^{2 \beta_0}
+ z^2 e^{-2 \chi} d_t \chi_s \ , \\
\alpha &=& z^2 \alpha_s \  , \; \; \;
\theta = z^2 \theta_s \ , \; \; \;
d_t \alpha = z d_t \alpha_s \ , \; \; \;
d_t \theta = z d_t \theta_s \ .
\end{eqnarray*}
The definition of the subscripted $s$ functions while somewhat arbitrary is guided by some underlying principles.
At a minimum, we are required to subtract singular terms from the metric functions so that the boundary conditions
are well behaved at $z=0$.  It is then convenient to rescale the subtracted metric functions by powers of $z$ such
that the stress tensor can be extracted without trying to compute a high order numerical derivative of the solutions.

We then numerically integrate to find the subscripted $s$ functions.
We impose the following boundary conditions on these functions at the singular point $z=0$:
\begin{eqnarray}
\partial_z \beta_s &=& 0 \ , \\
\pi^A_s &=& 3 e^{-2 \beta_0} U_3^A \ , \\
\partial_z U_s^A &=& U^A_3 \ , \\
d_t \alpha_s &=& 0 \ , \; \; \;
d_t \theta_s = 0 \ .
\end{eqnarray}
In integrating the $d_t\chi$ equation, we are faced with a choice.  We can either apply the Dirichlet condition
\[
d_t \chi_s = \frac{1}{2} V_3 - \frac{3}{4} e^{2 \beta_0} \chi_3 \ , \\
\]
at $z=0$ or the apparent horizon Dirichlet condition (\ref{apphorcond}) at $z=1$.  We choose the latter as it allows us not to propagate the boundary value of $V_3$ forward in time.
In integrating the $U^A_s$ and $\beta_s$ equations, we were also faced with a choice.  We could have applied Dirichlet conditions $U^A_s = 0$ and $\beta_s=0$ at $z=0$ instead.  However, we find in general that in discretizing the differential operator that we need then to invert, Neumann boundary conditions produce matrices with a lower condition number.

To have a well defined Cauchy problem, we also need to give initial conditions.  In this case, a set of good Cauchy data is provided by bulk data for $\alpha$ and $\theta$ and boundary data for $\chi_3$ and $U^A$.  As our choice of initial conditions will be guided by a particular hydrodynamics problem, let us postpone a discussion until after we have reviewed some facts about relativistic conformal hydrodynamics.

We relegate to appendix \ref{sec:gorydetails} the precise numerical details of our algorithm.

\section{Numerical Relativistic Conformal Hydrodynamics}
\label{sec:numericalhydro}

Thermal field theories generically admit a hydrodynamic description of their long wavelength, low frequency modes provided the wavelengths are long compared to the mean free path of the particles and the frequencies are small compared to inverse scattering times.  In particular, field theories with gravity duals admit such a description.  The goal of this section is to write down hydrodynamic equations whose numerical solutions can be compared with the numerical solutions of the gravity model in the same low frequency, long wavelength regime.

In our case, the field theory dual to our gravity model is both relativistic and conformal which puts some additional constraints on the constitutive relations for the stress tensor. The only scale in our field theory is the temperature $T$.  Thus a hydrodynamic description will be valid when the typical wavelength $\lambda \gg 1/T$ and the typical frequency $\omega \ll T$.  
We follow \cite{Baier:2007ix} in our description.  We assume the stress tensor has the form
\begin{eqnarray}
T^{\mu\nu} = (\epsilon + p) u^\mu u^\nu +p  g^{\mu\nu} + \Pi^{\mu\nu}
\end{eqnarray}
where we define $\Pi^{\mu\nu}$ recursively\footnote{%
The implicit definition of $\Pi^{\mu\nu}$ makes our formulation of the hydrodynamic equations Israel-Stewart like.  Formally, higher than second order gradient corrections are present in the definition of $\Pi^{\mu\nu}$.  However, if one wanted a third or higher order accurate  formulation, additional terms should be added to the definition of $\Pi^{\mu\nu}$.
Note some second order terms are necessary in order for stability of the numerics.
}
 in a gradient expansion
\begin{eqnarray}
\Pi^{\mu\nu} &=& - \eta \sigma^{\mu\nu} - \tau_\Pi \left[ (D \Pi)^{\langle \mu\nu \rangle} + \frac{d+1}{d} \Pi^{\mu\nu} (\nabla \cdot u) \right] \nonumber \\
&& + \kappa \left[ R^{\langle \mu\nu \rangle} - (d-1) u_\alpha R^{\alpha \langle \mu\nu \rangle \beta} u_\beta \right] \nonumber \\
&&
+ \frac{\lambda_1}{\eta^2} {\Pi^{\langle \mu}}_\alpha \Pi^{\nu \rangle \alpha} - \frac{\lambda_2}{\eta} {\Pi^{\langle \mu}}_{\alpha} \Omega^{\nu \rangle \alpha} + \lambda_3 {\Omega^{\langle \mu}}_\alpha \Omega^{\nu\rangle \alpha} \ .
\end{eqnarray}
Conformality implies tracelessness of $T^{\mu\nu}$ which in turn yields a relationship  $\epsilon = d \, p$ between the energy density $\epsilon$ and pressure $p$ in $d$ spatial dimensions.  
To unpack these expressions, we need a number of subsidiary definitions.  
We are working with a metric with mostly plus signature.  
The four velocity $u^\mu$ has norm $u^\alpha u_\alpha = -1$.  
(In the fluid rest frame in Minkowski space $u^\mu = (1,0, \ldots, 0)$.)
The derivative $D \equiv u^\mu \nabla_\mu$.  The vorticity is
\begin{eqnarray}
\Omega^{\mu\nu} \equiv \frac{1}{2} \Delta^{\mu \alpha} \Delta^{\nu \beta} (\nabla_\alpha u_\beta - \nabla_\beta u_\alpha) \ ,
\end{eqnarray}
where we have defined a projector onto a subspace orthogonal to the four velocity:
\begin{eqnarray}
\Delta^{\mu\nu} \equiv g^{\mu\nu} + u^\mu u^\nu \ .
\end{eqnarray}
The shear stress tensor is
\begin{eqnarray}
\sigma^{\mu\nu} \equiv 2 \nabla^{\langle \mu} u^{\nu \rangle} \ .
\end{eqnarray}
The angular brackets $\langle \rangle$ on the indices indicate projection onto traceless tensors orthogonal to the four velocity:
\begin{eqnarray}
A^{\langle \mu \nu \rangle} \equiv \frac{1}{2} \Delta^{\mu\alpha} \Delta^{\nu \beta}(A_{\alpha \beta} + A_{\beta \alpha}) - \frac{1}{d} \Delta^{\mu\nu} \Delta^{\alpha \beta} A_{\alpha \beta} \ .
\end{eqnarray}
Note that with these definitions, both $\Pi^{\mu\nu}$ and $\Omega^{\mu\nu}$ are traceless and orthogonal to the four velocity
\[
u_\mu \Pi^{\mu\nu} = u_\mu \Omega^{\mu\nu} = 0 \ , \; \; \;
\Omega^\mu_\mu = \Pi^\mu_\mu = 0 \ .
\]

In 2+1 space time dimensions (even in curved space), it can be shown that the coefficients of $\lambda_1$ and $\lambda_3$ vanish.  Thus, we are left with the four transport coefficients $\eta$, $\tau_\Pi$, $\kappa$, and $\lambda_2$.
We will be interested in what follows in a metric of the form $ds^2 = -g(t,x,y) dt^2 + dx^2 + dy^2$.  In this case, the Weyl curvature vanishes and we can forget about $\kappa$ as well.
For a fluid dual to pure Einstein gravity in 3+1 dimensions, 
$\eta$ was first computed in \cite{Herzog:2002fn}, $\tau_\Pi$ in \cite{Natsuume:2007ty}, and $\lambda_2$ 
in \cite{VanRaamsdonk:2008fp}.
Assuming a normalization of the energy density where
\begin{eqnarray}
\epsilon = \left(\frac{4 \pi T}{3} \right)^3 \ , \; \; \;
p = \frac{1}{2} \left( \frac{4\pi T}{3} \right)^3 \ 
\label{EPnorm}
\end{eqnarray}
(such that $\epsilon -2 p = 0$),
we have
\begin{eqnarray*}
\label{etachoice}
\eta &=& \frac{1}{2} \left( \frac{4 \pi T}{3} \right)^2 \ , \\
 \eta \tau_\Pi &=& \frac{1}{36} \left( \frac{4 \pi T}{3} \right) \left( \sqrt{3} \pi - 9 \log 3 + 18\right) \ , \\
\lambda_2 &=& \frac{1}{36} \left(\frac{4 \pi T}{3} \right) \left( \sqrt{3} \pi - 9 \log 3 \right) \ .
\end{eqnarray*}

For the numerics, we will use the relations $\Pi^\mu_\mu=0$ and $u_\mu \Pi^{\mu\nu}=0$ 
to eliminate all of the components of $\Pi^{\mu\nu}$ except for $\Pi^{xx} - \Pi^{yy} \equiv B$ and $\Pi^{xy}$.  
The three conservation conditions $\nabla_\mu T^{\mu\nu} = 0$ along with the recursive definitions of $B$ and $\Pi^{xy}$
become the five differential equations used to propagate the five variables $T$, $u^1$, $u^2$, $B$, and $\Pi^{xy}$.
A similar numerical scheme was described and implemented in ref.\ \cite{Luzum:2008cw}.

\subsection*{\it 3.1 Boosted Black Brane Background with $g_{tt}$ and general $\chi$}

The relationship between hydrodynamics and gravity can be made arbitrarily precise \cite{Bhattacharyya:2008jc} in the sense of an asymptotic series.
In principle, a solution to the hydrodynamic equations of motion at a given order in the gradient expansion can be integrated
in the radial direction of the gravity spacetime to provide a solution to Einstein's equations, accurate at the same order in a gradient expansion.  Here, we review how to perform this matching at the zeroth level in the gradient expansion -- for ideal hydrodynamics where the effects of viscosity and the other transport coefficients can be ignored.  
 In practice, this matching is important for us as it allows us to choose the same initial conditions in our gravity and hydrodynamic simulations.

The matching makes use of the boosted black brane metric.
This metric has a dual interpretation as a fluid moving with four velocity $u^\mu$ and temperature $T$.
The line element for the boosted black brane metric is 
\begin{eqnarray}
ds^2 &=&  \frac{1}{r^2} \left[ 2u_\mu \, dx^\mu dr - f(b r) u_\mu u_\nu \, dx^\mu dx^\nu + P_{\mu\nu} \, dx^\mu dx^\nu \right] \ , \\
P_{\mu\nu} &=& g_{\mu\nu}+ u_\mu u_\nu \ , \; \; \; g_{\mu\nu} =  \eta_{\mu\nu}  + (1-g)\delta_{0 \mu} \delta_{0 \nu} \ , \\
f(r) &=& 1-r^3 \ .
\end{eqnarray}
The temperature is given by the relation
 $T =3 b/4 \pi$.
 The horizon is at $r = 1/b$; the conformal boundary at $r=0$.  When $b$ and $u^\mu$ are constant, this solution is an exact solution of Einstein's equations.  More generally, the metric will only be accurate at zeroth order in gradients.  Note that we have allowed for 
 an arbitrary $g_{tt} = -g$.

We would like to put this metric into our null coordinate system.  To that end, we consider a coordinate transformation
$x^\mu = y^\mu + \xi^\mu(r)$ where for the moment we do not alter $r$.  
The vector $\xi$ is chosen such that the $g_{\mu r}$ components of the transformed metric vanish.
The line element of the transformed metric takes the form
\begin{eqnarray}
ds^2 = \frac{1}{r^2} \left[ \left( g_{\mu\nu} + (b r)^3 u_\mu u_\nu  \right) dy^\mu dy^\nu - \frac{2 g^{1/2} }{\sqrt{f(b r) + (u^0)^2 (b r)^3 g }} \, dt \, dr  \right] \ .
\end{eqnarray}

In our coordinate system, we have specified the functional form of the spatial determinant:
\begin{eqnarray}
\label{detrel}
\frac{e^{4 \chi(z)}}{z^4} = \frac{1}{r^4} \left(f(b r) -  (b r)^3 u^0 u_0 \right) \ ,
\end{eqnarray}
which defines $r$ as a function of $z$.  In the $z$ coordinate system, we place the horizon at $z=1$ which relates $\chi(1)$ to $b$ and $u^0$:
\begin{eqnarray}
e^{2\chi(1)} = b^2 u^0 g^{1/2} \ .
\end{eqnarray}

We can reconstruct $\beta$ from the $g_{tr}$ component of the transformed metric and the derivative $dr/dz$:
\begin{eqnarray}
e^{2 \beta} = \frac{\sqrt{g}}{1 + u^A u_A (br)^3} \frac{z^2}{r^2} \frac{dr}{dz} \ .
\end{eqnarray}
The other defining functions of the transformed metric are straightforward to reconstruct
\begin{eqnarray}
e^\alpha &=& \sqrt{\frac{1+(u^1)^2 (br)^3}{1+(u^2)^2 (br)^3} } 
%
\; , \; \; \;
\sinh \theta = u^1 u^2 e^{-2 \chi(z)} b^3 r z^2 \ , \\
U^A &=& -\frac{u^A u_0  (b r)^3}{1 + u^A u_A (br)^3} 
%
\; , \; \; \;
V = 
\frac{\sqrt{g} f(br)}{\sqrt{1 + u^A u_A (br)^3}} \frac{1}{z^3} \frac{dz}{dr} \ .
\end{eqnarray}

The relation (\ref{detrel}) suggests a natural choice for $\chi$, namely
\begin{eqnarray}
\chi = \frac{1}{4} \log (1 + u^A u_A z^3) + \log b \ .
\end{eqnarray}
Allowing us to make the identifications
\begin{eqnarray}
rb = z \ , \; \; \;
2 \chi_3 = u^A u_A \ .
\end{eqnarray}
With our gauge choice for $\chi$, these relations reduce to the simpler
\begin{eqnarray}
e^{2 \beta} &=& \frac{b \sqrt{g}}{\sqrt{1+ u_A u^A z^3}}  \ , \\
e^{2\alpha} &=& \frac{1+(u^1)^2 z^3}{1+(u^2)^2 z^3} 
\; , \; \; \;
\sinh \theta = \frac{u^1 u^2 z^3}{\sqrt{1 + u^A u_A z^3} } \ , \\
U^A &=& -\frac{u^A u_0 z^3}{1+ u^A u_A z^3} 
\; , \; \; \;
V = \frac{ b f(z) \sqrt{g}}{z^3 \sqrt{1 + u^A u_A z^3}}\ .
\end{eqnarray}
Seting $b=1$ recovers the gauge choice (\ref{chitest}) described above.

\section{A Simple Experiment}
\label{sec:results}

For our simple experiment, we start with a constant fluid flow in the $x$ direction at time $t=0$ and time-time component of the metric of the form 
\begin{eqnarray}g_{tt} = -(1 + \delta \cos (kx) e^{-m/t}). \label{gttbdy} \end{eqnarray}
 At time $t=0$, the metric reduces to Minkowski space while for $t \gg m$,
the fluid experiences a roughly constant sinusoidal potential in the $x$-direction.  As the potential breaks translation invariance,
we expect that the fluid velocity will eventually relax to zero.
A nonzero $m$ is used solely to increase the stability of the numerical simulations.  In our analytic estimates, we assume $m=0$.  We expect these estimates to be valid for times $t \gg m$.  

\subsection{Analytic Estimates of Momentum Relaxation}

Before entering a discussion of our numerical simulations, let us begin with three analytic estimates of the momentum relaxation rate.  The first is valid in the regime where hydrodynamics and linear response are both valid, the second when linear response is to be trusted, and the third when hydrodynamics is valid.

The technique we shall use in both the first and second cases  is called the memory function formalism and 
relies on the validity of linear response.  In other words, the metric source must be small $\delta \ll 1$.  
(Although not necessary, we will also assume the fluid velocity is small. We relax this assumption in appendix B.)  
This method was first used in a holographic context by ref.\
\cite{Hartnoll:2007ih}.   Later uses include refs.\ \cite{Hartnoll:2008hs,Hartnoll:2012rj}.
The method is described in detail in the book \cite{Forster}.

In this method, we break translation invariance by adding the following perturbation to the action
\begin{eqnarray}
 \delta \int \calo(x) e^{ikx} dx \ ,
\end{eqnarray}
where $\calo(x)$ is an arbitrary operator.
The memory function formalism can be used to deduce the relaxation time:
\begin{eqnarray}
\frac{1}{\tau} =  \frac{\delta^2 k^2}{\epsilon + p} \left\{ \left. \lim_{\omega \to 0} \frac{ \operatorname{Im} G^{\calo \calo}_R (\omega ,k) }{\omega} \right|_{\delta=0} \right\} \ ,
\end{eqnarray}
where $G^{\calo \calo}_R(\omega ,k)$ is the Fourier transform of the retarded Green's function.  
The method is perturbative, and one may worry about the running of $\delta$ with energy scale.  
In our experiment, in order for this method
to be reliable, we need $\delta$ to remain small at the scale set by the temperature $T$.  (In the limit $T \to 0$, this restriction would mean that $\calo(x)$ should be marginal or irrelevant.)
In our case, we are perturbing the metric component $g_{tt}$ which couples to the energy density $T^{tt}/2$, 
a marginal operator.\footnote{%
 The factor of two comes from the canonical definition of the stress tensor.
 }
Thus we need to compute the   
retarded Green's function for 
the energy density $G^{\epsilon \epsilon}_R(\omega, k)$.

\subsubsection*{\it 1. Hydrodynamic regime} 

We first compute this Green's function purely in the hydrodynamic limit where $k \ll T$.  
This Green's function has a universal form (see for example \cite{Kovtun:2012rj}):
\begin{eqnarray}
G^{\epsilon \epsilon}_R = \frac{k^2 (\epsilon + p)}{ k^2 (c_s^2 + i \Gamma \omega) - \omega^2} + \epsilon \ ,
\end{eqnarray}
where $c_s^2 = \partial p / \partial \epsilon$ is the speed of sound squared while the damping constant is 
\[
\Gamma = \frac{1}{\epsilon + p} \left( \frac{2(d-1)}{d} \eta + \xi \right) \ .
\]
As it is simple to do, we have kept the spatial dimension $d$ 
arbitrary and restored the bulk viscosity $\xi$, which will not be present for a 
conformal fluid.
Plugging this universal form into the expression for the relaxation time yields
\begin{eqnarray}
\frac{1}{\tau} = \frac{\delta^2 k^2}{4 sT c_s^4}  \left( \frac{2(d-1)}{d} \eta + \xi \right) \ .
\label{linresp}
\end{eqnarray}
The speed of sound appearing in the denominator implies that this relaxation effect will disappear in the incompressible limit where $c_s \to \infty$ and sound waves can be neglected.
For a cosine perturbation instead of the $e^{ikx}$ dependence considered here, we multiply the result by a factor of one half.

\subsubsection*{\it 2. Short wavelength limit}

Next we compute the Green's function directly from gravity.  The result should be applicable away from the hydrodynamic limit, where $k$ is no longer necessarily small.
We follow the gauge invariant formulation of ref.\ \cite{Kovtun:2005ev} where a corresponding calculation was performed for a five dimensional spacetime.  We begin instead with the black brane metric for a four dimensional space-time:
\[
ds^2 = \frac{1}{z^2} \left( -f(z) dt^2 + dx^2 + dy^2 + \frac{dz^2}{f(z)} \right) \ ,
\]
where $f(z) = 1-z^3$.  Thus, the horizon of our black hole is at $z=1$ and we will be measuring everything in units of the horizon radius.  (At the end of the day, we can restore the temperature dependence by performing the rescalings $\omega \to 3 \omega / 4 \pi T$ and $k \to 3 k / 4 \pi T$.)  
We consider small diffeomorphisms of the form $g_{\mu\nu} \to g_{\mu \nu} + \xi_{\mu;\nu} + \xi_{\nu;\mu}$ where we restrict
$\xi_\mu = \xi_\mu(z) e^{-i\omega t+ i k x}$.  We find that the following linear combination of metric fluctuations, again restricted to have the form $h_{\mu\nu} = h_{\mu\nu}(z) e^{-i \omega t + i k x}$, is invariant under such gauge transformations:
\begin{eqnarray}
Z = z^2 \left[ k^2 h_{tt} + 2 \omega k h_{tx}  + \omega^2 h_{xx} + \left( k^2 f(z) - \omega^2 - \frac{z}{2} k^2 f'(z) \right) h_{yy}\right] \ .
\end{eqnarray}
This gauge invariant combination satisfies the second order linear differential equation
\begin{eqnarray}
\lefteqn{Z'' +\left( \frac{2}{z} -\frac{f'}{f} + \frac{6k^2 z^2}{h(\omega,k;z)} \right) Z' +}  \nonumber \\
&& \left( \frac{\omega^2}{f^2} - \frac{k^2}{f} + \frac{9k^2z}{(3 k^2-4 \omega^2)f} +\frac{36 k^2 (k^2-\omega^2)z}{(3k^2-4\omega^2)h(\omega,k;z)} \right) Z = 0 \ ,
\label{Zeq}
\end{eqnarray}
where we have defined $h(\omega,k;z) \equiv k^2(-4+z^3) + 4\omega^2$.

To deduce the retarded Green's function, at the horizon $z=1$, we enforce ingoing boundary conditions
\begin{eqnarray}
Z =  (1-z)^{-i \omega /3}\left( C(\omega, k) + O(1-z) \right)\ .
\end{eqnarray}
At the conformal boundary $z=0$, we find a series expansion of the solution
\begin{eqnarray}
Z(z) = A(\omega ,k) (1 + O(z^2)) + B(\omega,k) (z^3 + O(z^5)) \ .
\end{eqnarray}
Up to real contact terms, the retarded Green's function can then be extracted from the ratio $B / A$. 
There is an overall normalization missing from this Green's function which we can choose to gain agreement with 
the hydrodynamic result (\ref{linresp}) along with the choice (\ref{EPnorm}).  We will see that in units where $T = 3/4 \pi$, we need 
\begin{eqnarray}
G_R^{\epsilon \epsilon}(\omega, k) = \frac{3}{2} \frac{B(\omega, k)}{A(\omega, k)} \ .  
\end{eqnarray}

We are interested in only the imaginary part of the Green's function for which it is enough to know the absolute values
$|A|$ and $|C|$.  To see why, it simplifies the algebra to put the differential equation in Schr\"odinger form.  
We define a new wave function $\psi$ such that
\begin{eqnarray}
Z = \frac{z}{\sqrt{f}} (4 \omega^2 + k^2 (-4 + z^3)) \psi \ ,
\end{eqnarray}
In this case, the differential equation reduces to $- \psi'' + V \psi = 0$
where the potential is given by
\begin{eqnarray}
V ={2\over z^2}+{k^2 \left( 18 z^4 + h(\omega, k;z) \right) \over (1-z^3) h(\omega, k;z)} +{18k^4 z^4 \over h(\omega, k;z)^2} -  {4 z^2 \omega^2 + 9z^6 \over 4z^2(1-z^3)^2}  \ .
\end{eqnarray}
We now find boundary behaviors
\begin{eqnarray}
\psi(z) = \left( \frac{1}{z} + \ldots \right) \tilde A +  (z^2 + \ldots)\tilde B \; \; \mbox{and} \; \; 
\psi(z) \sim  (1-z)^{-i \omega/3+1/2} \tilde C \ 
\end{eqnarray}
at $z=0$ and $z=1$ respectively.  The ratio $\tilde B/\tilde A= B/A$ is invariant.  
Given complex conjugate solutions $\psi$ and $\psi^*$ for real $\omega$, their corresponding Wronskian must be constant.  In other words, there is a conserved probability current.  This conservation condition implies
\begin{eqnarray}
\operatorname{Im} \frac{ B}{ A} = \frac{|\tilde C|^2}{| \tilde A|^2} \frac{\omega}{9} \ .
\end{eqnarray}

\begin{figure}[t]
   \centering
  \includegraphics[scale=0.65]{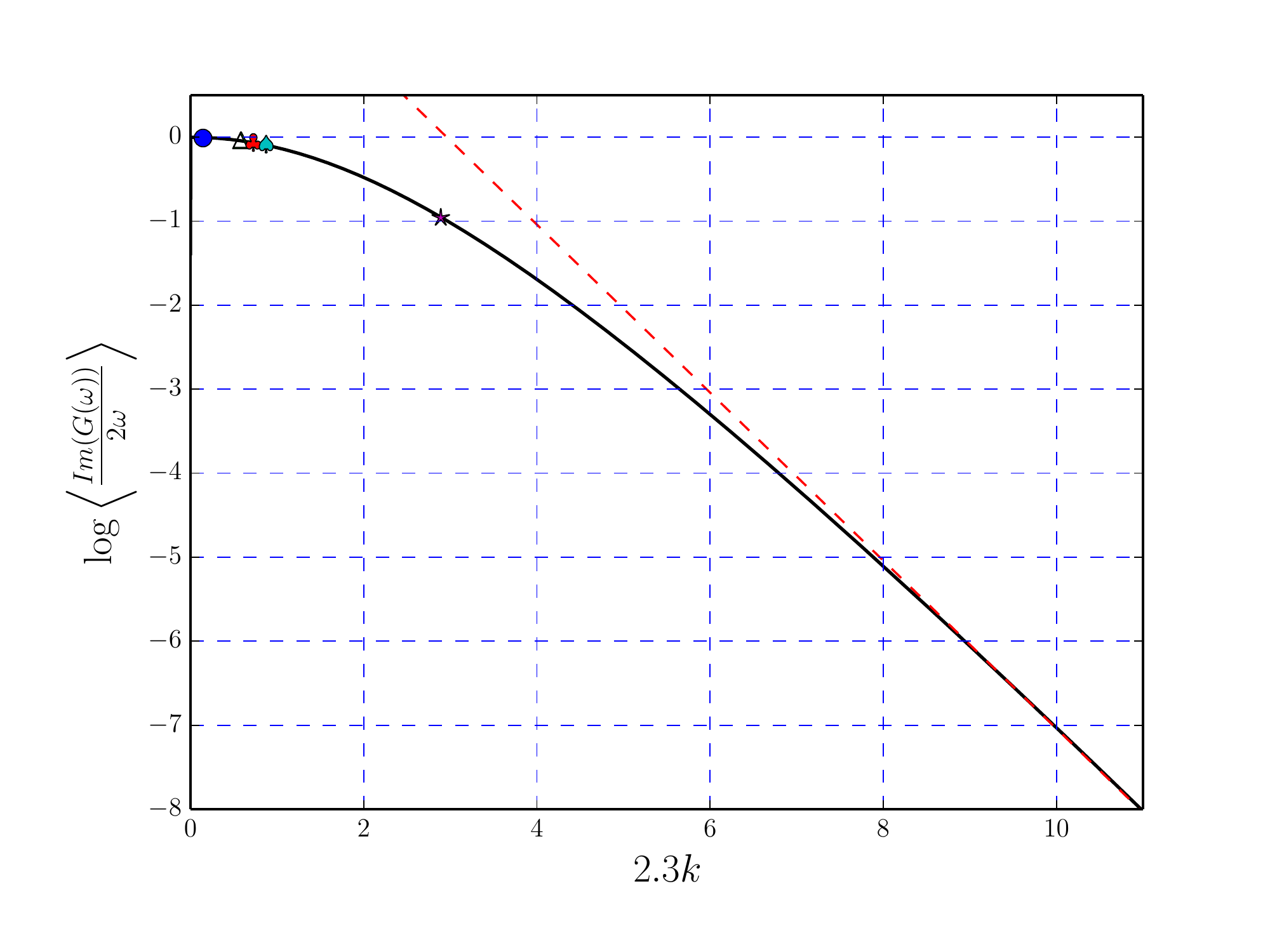} 
     \caption{  The variation of 
     $\displaystyle{\mathop{\lim}_{\omega \to 0}} {\operatorname{Im} G(\omega) \over 2 \omega}$ 
     as a function of the lattice wave number. 
     The black curve shows the result obtained from solving (\protect{\ref{Zeq}}) numerically, i.e.\ linearized gravity.
     The dotted red line shows the approximate behavior of the Green's function for large values of $k$.  
The markers $\bullet$, $\Delta$, $\clubsuit$,  $\spadesuit$ and $\star$ show the values obtained  by solving the full nonlinear gravity equations for $k=\pi/50 ,~ 4\pi/50 ,~ 5\pi/50 ,~ 6\pi/50 $ and $20\pi/50$ in the linearized regime 
($\delta=0.2$). Note that $k$ is expressed in units where $3/(4\pi T)=1$.
The $x$-axis is rescaled so that the red line has slope minus one.}
   \label{fig:imG}
\end{figure}

Figure \ref{fig:imG} is a numerical determination $ \lim_{\omega \to 0} \operatorname{Im} G_R^{\epsilon \epsilon} (\omega, k) /2 \omega$.  For small $k$, the value is very close to 1 while for large $k$, the Green's function is exponentially damped.  We can recover these two limits analytically.
For small $k$, the differential equation can be solved in a hydrodynamic expansion where $k,\omega \ll 1$.  The answer is 
\begin{eqnarray}
Z(z) = C f(z)^{-i \omega / 3} \left(1 + \frac{k^2 f(z)}{4 \omega^2 - 3k^2} - \frac{4 i \omega}{3} + \ldots \right) \ .
\end{eqnarray}
Note the leading $f(z)^{-i \omega / 3}$ factor enforces ingoing boundary conditions at the horizon $z=1$.
From this answer it follows that
\begin{eqnarray}
\lim_{k \to 0} \lim_{\omega \to 0} \operatorname{Im} \frac{B(\omega,k)}{\omega A(\omega, k) } = \frac{4}{3} \ .
\end{eqnarray}

Next, we can solve this differential equation in a WKB limit when $k \gg 1$.
As described in \cite{Son:2002sd}, the imaginary part of the stress tensor Green's function is given by the tunneling probability through the potential $V$.  In the large $k$ limit, this potential reduces to $k^2/f$.
Taking $\tilde A = 1$, the WKB connection formulae give that 
\begin{eqnarray}
\tilde C \sim \exp \left(- k \int_0^1  \frac{dz}{\sqrt{f(z)}} \right) = \exp \left( -k \sqrt{\pi} \frac{\Gamma(4/3)}{\Gamma(5/6)} \right) \ .
\label{WKBscaling}
\end{eqnarray}
Numerically, the factor multiplying $k$ in the exponent is approximately 1.40.  
The imaginary part of the energy density Green's function should then scale as $|\tilde C|^2$.  
In this WKB analysis, we have swept under the rug subtleties associated with the quadratic singularities in the potential at $z=0$ and $z=1$ and also the fact that there is a classical turning point in the potential at $z_0<1$ an $O(k^2)$ distance from $z=1$.  These subtleties produce $1/k^2$ corrections to the tunneling amplitude.
In our numerics, we have not been able to get to large enough values of $k$ to see the scaling (\ref{WKBscaling}) although we do see exponentially damped behavior with a slightly smaller exponent (2.3 instead of 2.80) in Figure \ref{fig:imG}.

 \subsubsection*{\it 3. Small fluid velocity limit for arbitrary $\delta$}

When $\delta$ is large, it is not possible to use linear response theory to compute the relaxation time scale.
It is possible to obtain an expression for the relaxation time scale by studying linearized fluctuations around the
exact solution of second order hydrodynamics, to which the system must relax eventually.  We can check that the
following is a solution of the hydrodynamic equations\footnote{%
Interestingly, 
this same solution is used as a starting point to find the non-dissipative transport coefficients in a Lagrangian formulation of hydrodynamics in refs.\ \cite{Banerjee:2012iz,Jensen:2012jh}.
}
 (with $\eta \neq 0$):
\begin{eqnarray}
u^1 = u^2 = 0 \ ; \; \; \; T = \frac{T_0}{\sqrt{g(x)}} \ .
\label{Analytic}
\end{eqnarray}
We expect that for any arbitrary initial conditions, the relativistic flow will relax to the above solution in the presence
of viscosity. Note that the parameter $T_0$ appearing in the solution is related to the spatial average of the final steady state temperature as follows
\begin{eqnarray}{ \bar{T}_f} ={1\over L} \int_{-L/2}^{L/2} T dx = \left( {2 T_0\over \pi \sqrt{1-\delta} }\right)  K\left({-2\delta\over 1-\delta}\right) \label{Tfeq}\end{eqnarray}
where $K(\nu)$ is the quarter period of the Jacobi elliptic function and $\nu^2$ is the elliptic modulus. We will make use of this expression when we present the numerical results. 

In the limit where the flow velocity is small, the temperature and velocity profiles can be expanded as
\begin{eqnarray}
T &=&
\frac{T_0}{\sqrt{g(x)}} + \varepsilon \, T_1 (x,t) + O(\varepsilon^2) \ , \\
u^1 &=&
\varepsilon \, v_1(x,t) + O(\varepsilon^2) \ ,
\end{eqnarray}
where $\varepsilon$ is a small number parametrizing the speed of the flow. 

We now specialize to relativistic conformal hydrodynamics at first order in the gradient expansion and take the metric
source to be of the form $g = 1 + \delta \cos kx$.
The hydrodynamic equations linearize in the small flow velocity regime, allowing us to make a separation of variables ansatz $v(x,t) = e^{-\lambda k^2 t} v_1(kx)$
and $T_1(x,t) = e^{-\lambda k^2  t} T_1(kx)$.
The resulting pair of ordinary differential equations in $x$ are then
\begin{eqnarray}
0&=& v_1'(kx) + \frac{\delta \sin kx}{2(1+\delta \cos kx)}v_1(kx) - \frac{2 \lambda kT_1(kx)}{ T_0}  \ , \\
0 &=& v_1''(kx) - \frac{3 \epsilon_0}{2 \eta_0 k} T_1'(kx) + \frac{3 \epsilon_0 \delta \sin kx}{4 \eta_0 k(1+\delta \cos kx)} T_1(kx)
\\
&&+ \frac{12 \lambda \epsilon_0 T_0 - 5 \eta_0  \delta^2 - 4(\eta_0  - 3 \lambda \epsilon_0 T_0) \delta \cos kx
+ \eta_0  \delta^2 \cos 2kx}{8 \eta_0 (1+\delta \cos kx)^2} v_1(kx)
\ , \nonumber
\end{eqnarray}
where we have parametrized the energy density and viscosity as  $\epsilon = \epsilon_0 T^3$ and $\eta = \eta_0 T^2$ respectively.

We solve this pair of equations to leading order in $k$.
It is pointless to keep terms beyond leading order as we have already thrown out the second order corrections to the hydro equations.
The solution is
\begin{eqnarray}
v_1(y) &=& v_0 \sqrt{1+\delta \cos y}  + O(k^2)\ , \\
T_1(y) &=& k \frac{(3 \lambda \epsilon_0 T_0 - 2 \eta_0) y + 4 \eta_0 \sqrt{1-\delta^2} \tan^{-1} \left( \sqrt{\frac{1-\delta}{1+\delta}} \tan \frac{y}{2} \right) }{\sqrt{1+\delta \cos y}} + O(k^3)\ .
\end{eqnarray}
In order for $T_1$ to be a continuous function of $y$, we must have that
\begin{eqnarray}
\lambda = \frac{2 \eta_0}{3 \epsilon_0 T_0} \left( 1 - \sqrt{1-\delta^2} \right) \ .
\label{larged}
\end{eqnarray}
The momentum relaxation time scale is given by $\tau^{-1}=(\lambda k^2)$.
When $\delta>1$, the metric changes signature and hence the momentum relaxation time scale becomes complex.
In the limit where $\delta \ll 1$, we recover the result (\ref{linresp}) obtained using linear response theory. 

\subsection{Numerical Results}

In this subsection, we study momentum relaxation numerically in the presence of the metric source (\ref{gttbdy}) using (i) gravity equations in an asymptotically $AdS_4$ space-time and  (ii) second order hydrodynamics equations. 
The intial conditions for $\alpha$, $\theta$, $\chi$ and $U^A_3$ are obtained from the boosted black brane metric in \S 3.1 with $u^A = (0.2,0)$. In the following, we study the dependence of relaxation time on the lattice parameters $\delta$, $m$ and $k$ for a fixed box size $L= 100\times 3/(4\pi T_i) $ where $T_i$ is the initial temperature. In future discussions, $k$, $T$, $1/x$, $1/t$ and $m$ 
are expressed in units where $3/(4\pi T_i)=1$. 

\subsubsection*{\it 1. Relaxation time scale for large lattice spacing and small $\delta$}
\begin{figure}
   \centering
a)    \includegraphics[scale=0.65]{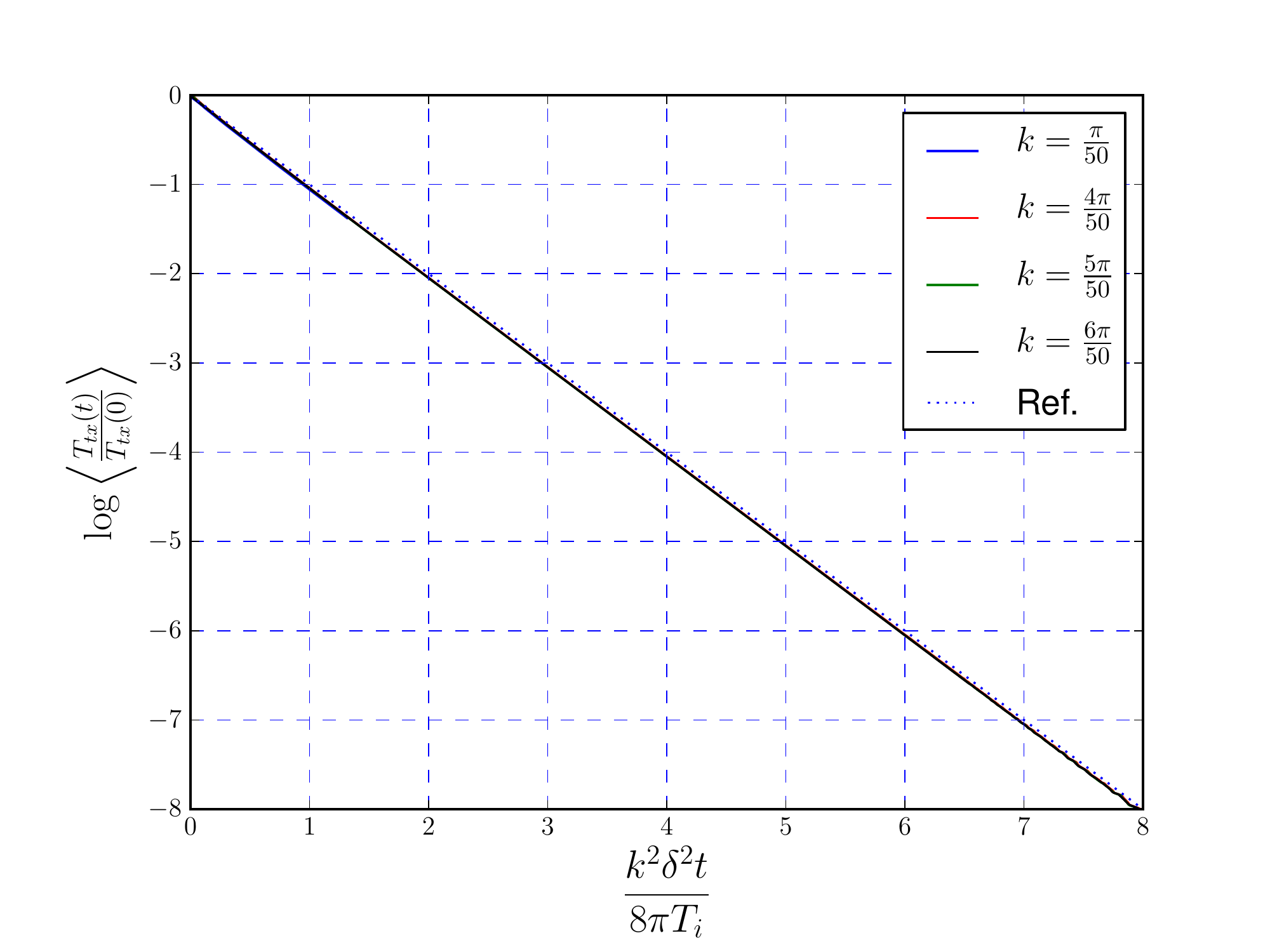} \\ 
  b)      \includegraphics[scale=0.65]{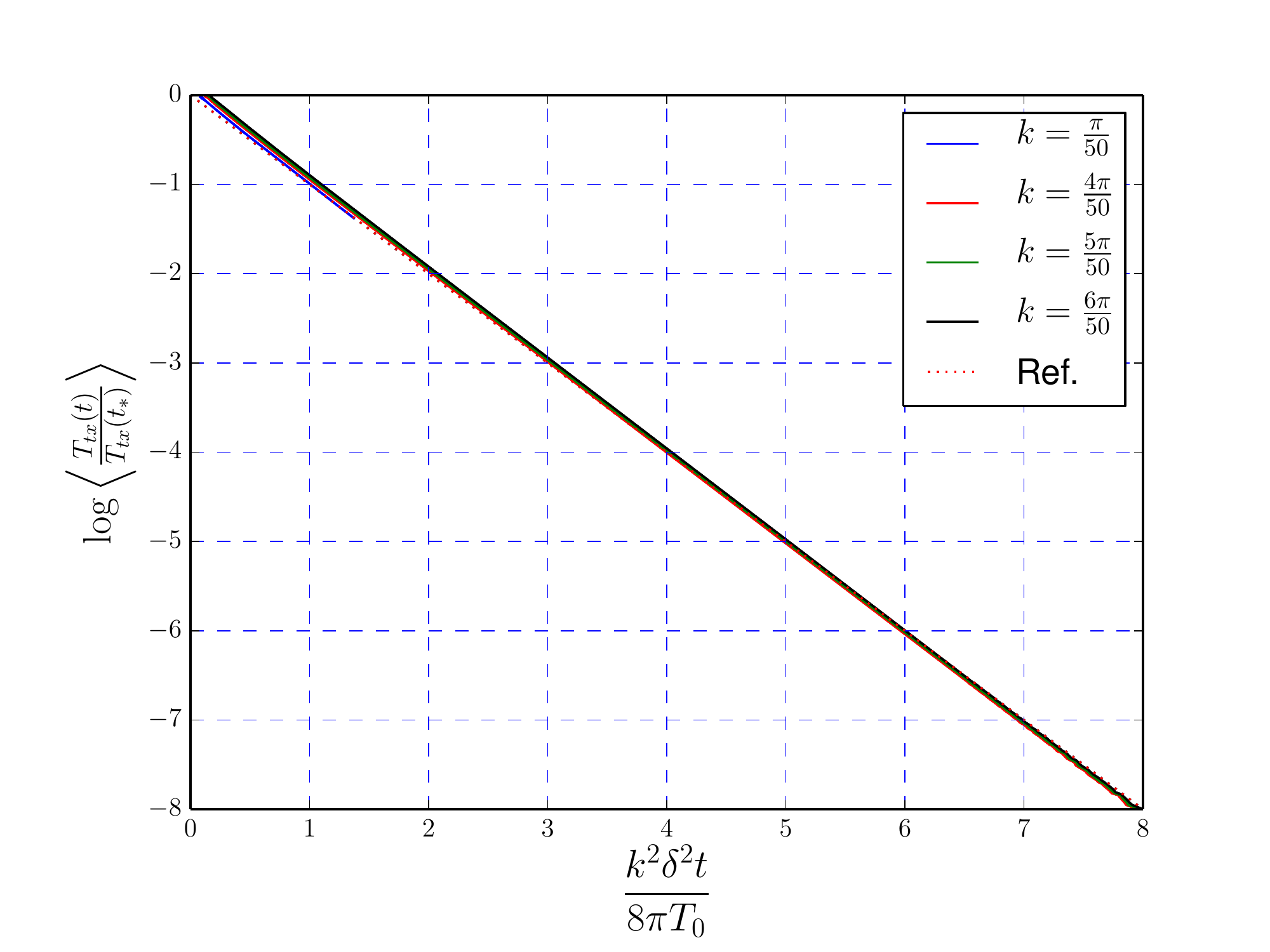} 
     \caption{A plot of $\log \langle T_{tx} \rangle$ as a function of scaled time for different values of $k$, obtained from hydrodynamic simulations. The analytical expression for relaxation time computed in the previous section corresponds to the reference line with slope -1.0.  In (b), values of $t_*$  are chosen such that the lines agree at late times.}
   \label{fig:hsmallk}
\end{figure}

\begin{figure}
   \centering
  a) \includegraphics[scale=0.65]{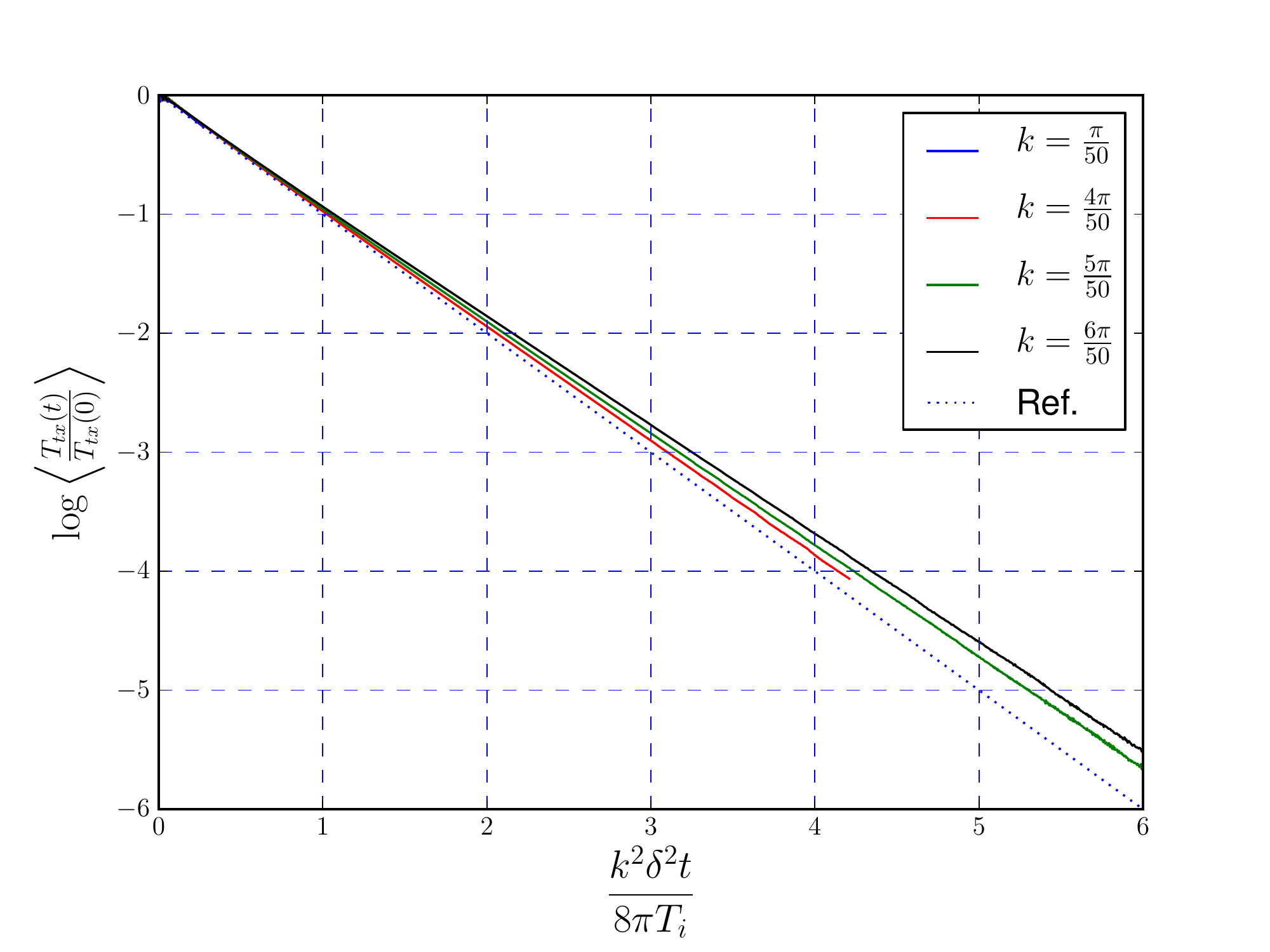} \\
   b)  \includegraphics[scale=0.65]{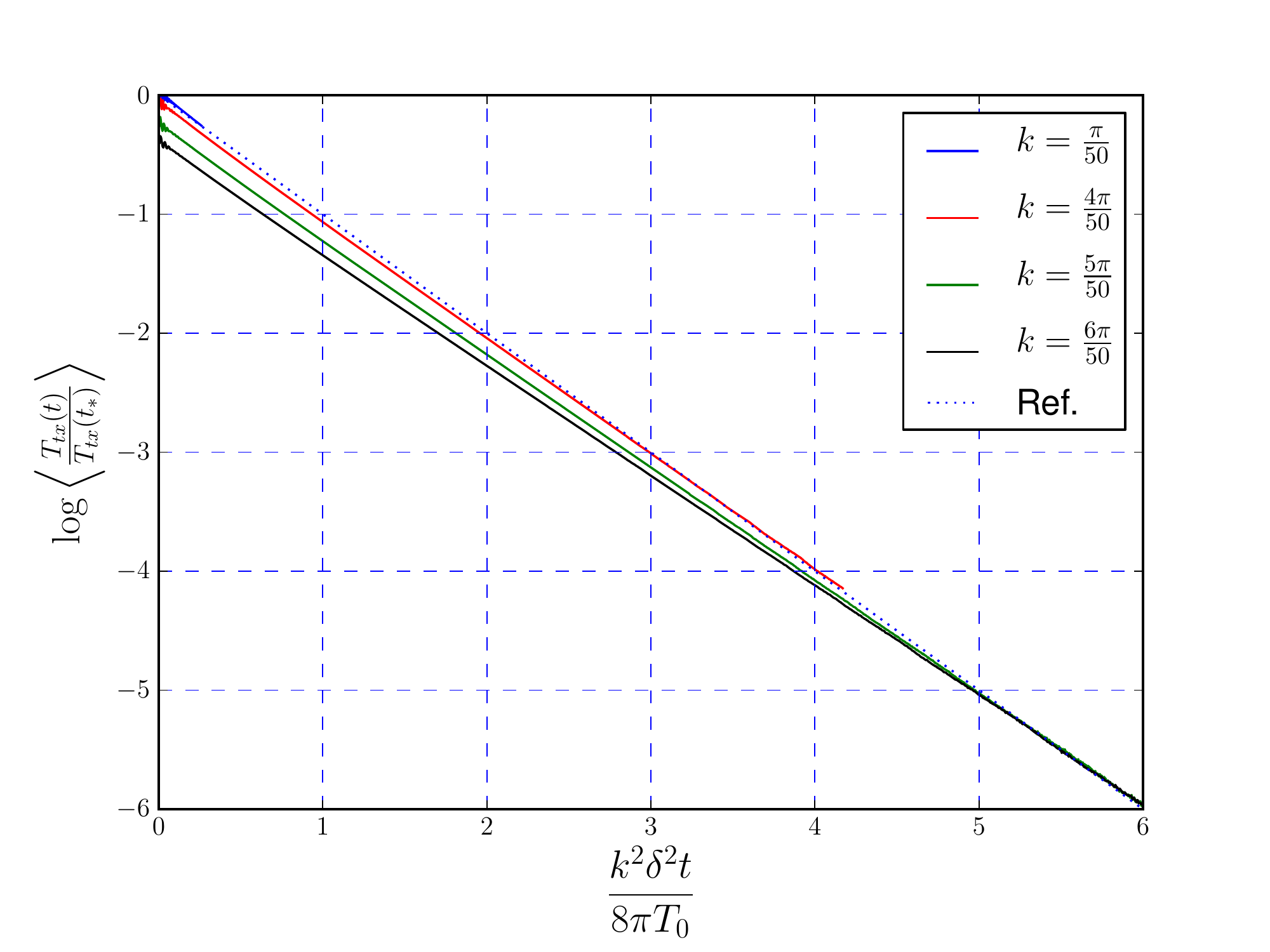} 
     \caption{  A plot of $\log \langle T_{tx} \rangle$ as a function of scaled time for different values of $k$, obtained from gravity simulations. The analytical expression for relaxation time computed in the previous section corresponds to the reference line with slope -1.0.  In (b), values of $t_*$ are chosen such that the lines agree at late times. The simulations were run for $25 \times 10^4$ time steps with $\Delta t = 0.05$.
     }
     \label{fig:grsmallk}
\end{figure}

First we will show that at small $k$, the results obtained from gravity and hydrodynamic simulations agree. Figs.\ \ref{fig:hsmallk}a  and  \ref{fig:grsmallk}a show plots of $\log \langle T_{tx} \rangle$ for different values of $k$ (with $ 3k/4 \pi T \ll 1$  ) obtained using hydrodynamics and gravity respectively. In these examples, we assume that $\delta = 0.2$, $m=20$, 
and the number of spatial grid points used for the simulation is $N = 101$. Note that the time is rescaled by a factor of $\eta \delta^2 k^2 / 8 \pi T_i$, which is the inverse relaxation time scale obtained analytically using linear response theory for hydrodynamics. $T_i$ is the initial temperature. The slope of the plot of $\log \langle T_{tx} \rangle$ is approximately -1.0 initially.  The reference line has slope -1.0 which is the linear response theory result. We conclude that the results of the full non-linear hydrodynamics and gravity simulations agree with the linear response theory.

However, the plots show deviations from the linear response theory at late times. The small deviations in the hydrodynamic computation arise from the slow variation of temperature (\ref{Tvariation}). (As we review in appendix B, energy conservation relates the decrease in flow velocity to an increase in temperature.) Now let us look at Fig.\ \ref{fig:hsmallk}b.  In this plot, time is scaled by a factor of $\eta \delta^2 k^2 / 8 \pi T_0$, where $T_0$ is computed from the final temperature using (\ref{Tfeq}). This rescaling allows us to see that at late times the full non-linear hydrodynamics agrees with the result in (\ref{larged}) when $\delta$ is small. 

In the gravity simulations, the final temperature is computed using the boundary stress tensor as follows. First we compute the energy density $\varepsilon$ as the eigenvalue associated with the time-like eigenvector of the boundary stress tensor: $\langle{T^{\mu\nu}}\rangle u_\nu = \varepsilon u^\mu$, where $u^\mu$ is time-like. We can then compute the  
temperature\footnote{%
The temperature can also be computed from the surface gravity of the apparent horizon. The surface gravity is given by $\kappa = l^\mu n^\nu \nabla_\nu n_\mu$ where $ l_\mu dx^\mu = dt$ and $n_\mu dx^\mu = dr$. This definition assumes that the apparent horizon is a Killing horizon. However, in non-static spacetimes the apparent horizon need not be a Killing horizon, leading to an ambiguity in the definition.
}
 assuming the equation of state $\epsilon = (4\pi T/3)^3$.

It is clear from Fig. \ref{fig:grsmallk}a that the deviations from hydrodynamic linear response theory at late times in the gravity simulations are larger than in the hydrodynamic simulations.  
As we did in the hydrodynamic plot, we can get better agreement at late times by rescaling by $T_0$ instead of $T_i$ (Fig.\ \ref{fig:grsmallk}b).  The deviations in gravity then become greater at early times.  
The deviations we are seeing come from gradient corrections to hydrodynamics.
We can infer from Fig.\ \ref{fig:imG} that the relaxation time computed using linear response theory in gravity differs from hydrodynamic linear response theory when $k$ is not close to zero. 
In fact, one can check that the relaxation time scales computed from Fig.\ \ref{fig:grsmallk}a agree with Fig.\ \ref{fig:imG}.

These results are treated as checks on our numerical results. We present additional checks on the numerics in appendix
\ref{sec:gorydetails}. In particular, we show that in the small $k$ limit, the stress tensors computed from hydrodynamic and gravity simulations are in good agreement.

\begin{figure}[h]
   \centering
    \includegraphics[height=300pt]{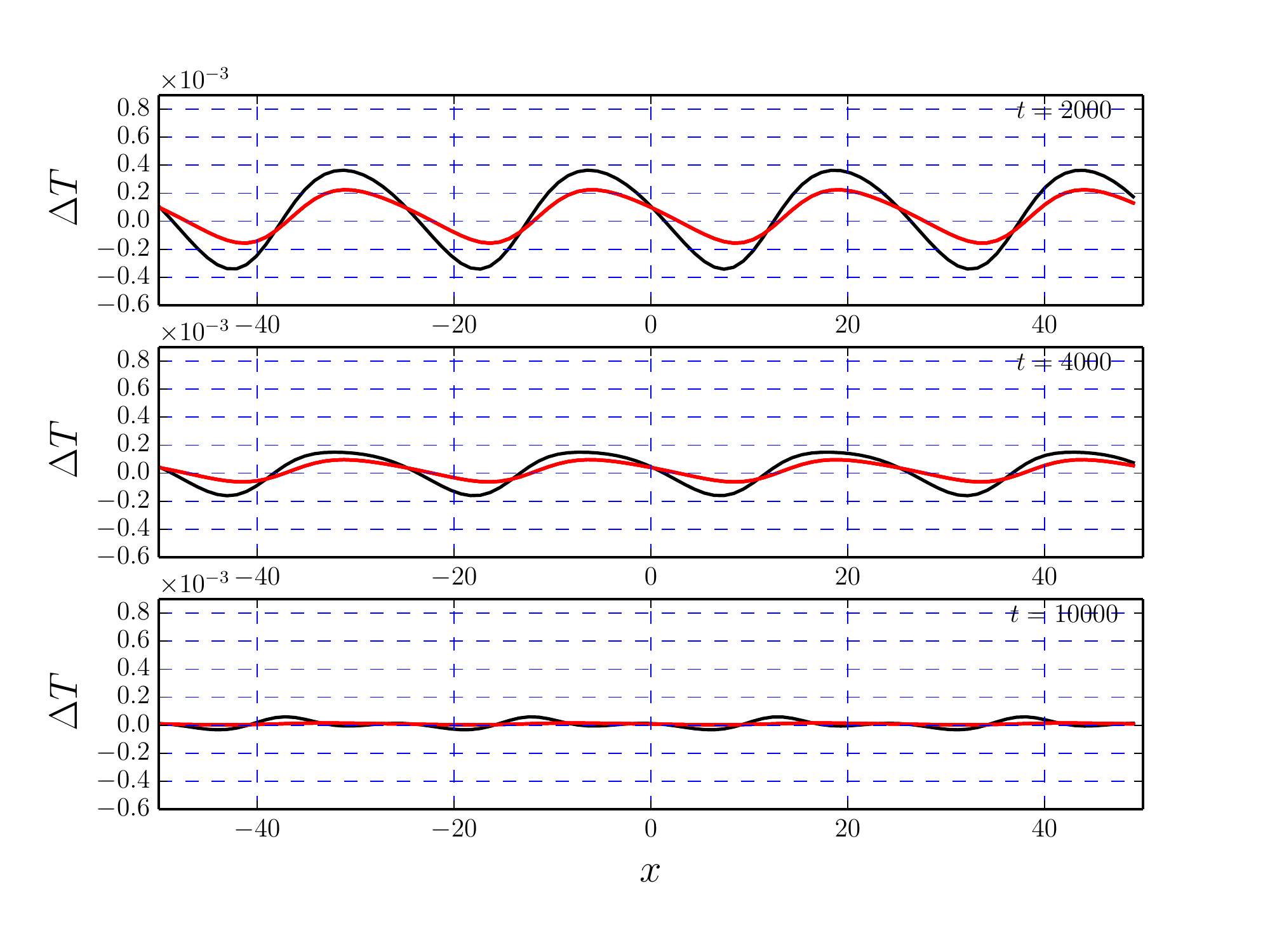} 
     \caption{ A plot of the difference in the numerically computed temperature and the exact analytical expression for temperature in (\ref{Analytic}) at $t=2000$, 4000 and 10{,}000. The lattice wavenumber is $k=4\pi/50$. The red curve corresponds to the result obtained from hydrodynamic simulations and the black curve corresponds to gravity simulations. We use the final value of mean temperature computed from gravity and hydrodynamic simulations to compute $T_0$.}
   \label{fig:temppr}
\end{figure}

At late times, the system approaches the equilibrium solution described in (\ref{Analytic}). Figs.\ \ref{fig:hsmallk}  and \ref{fig:grsmallk} show that the momentum relaxes to zero at late times. Furthermore, Fig.\ \ref{fig:temppr} shows that the temperature profile at late times agrees with the expression in (\ref{Analytic}). In particular, the difference between the numerically computed value of temperature and the analytic expression decreases as time evolves, and the gradients in $\Delta T$ become smaller.\footnote{%
 It is tempting to interpret the deviations of the gravity solution from (\ref{Analytic}) in the third panel of Fig.\ \ref{fig:temppr} as gradient corrections.  However, the deviations are comparable to our numerical error.
}

\subsubsection*{\it 2. Relaxation time scale for small lattice spacing and small $\delta$}

\begin{figure}[h]
   \centering
    \includegraphics[height=250pt]{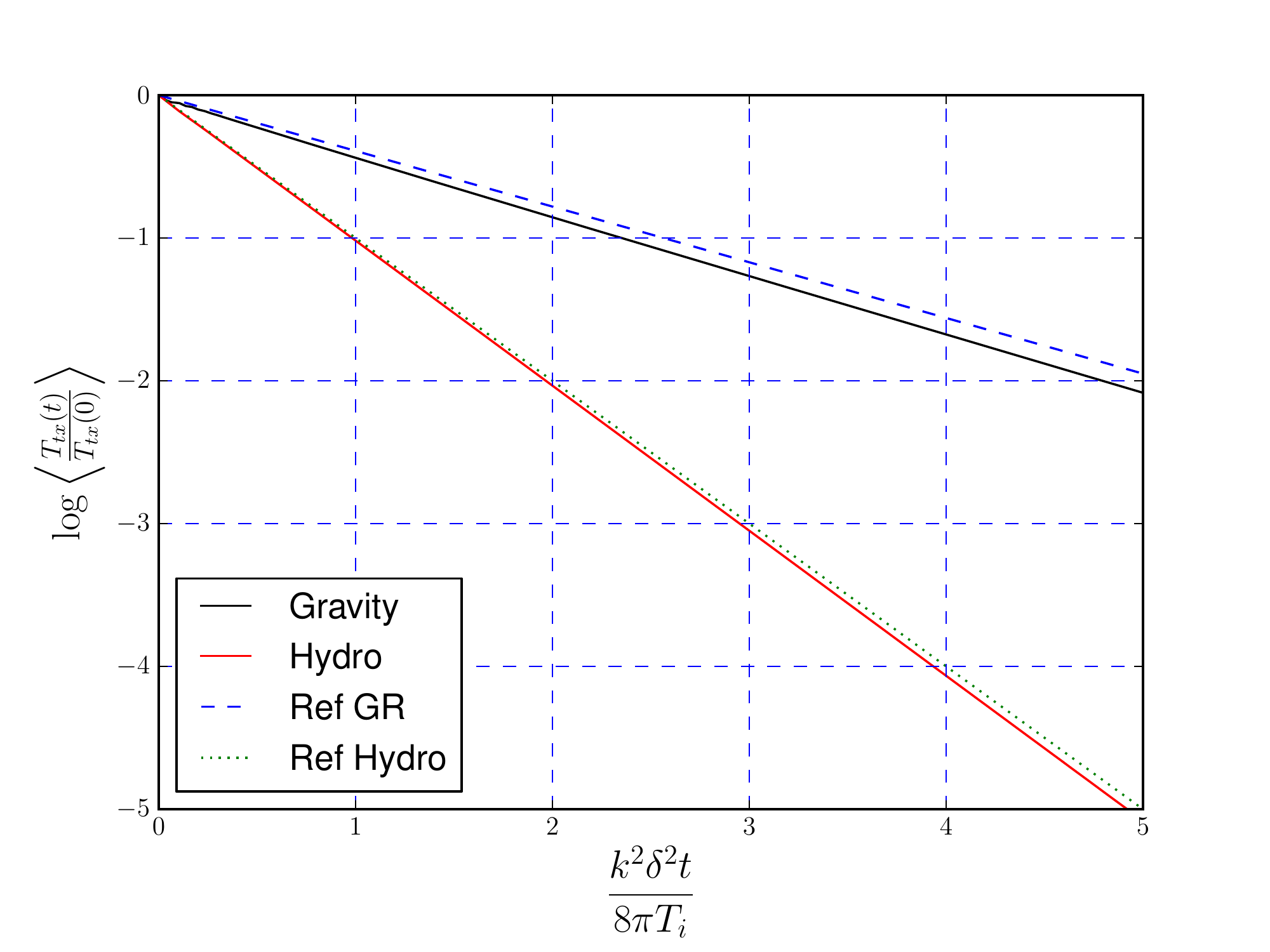} 
     \caption{ A plot of $\log \langle T_{tx} \rangle$ as a function of scaled time when $k=20 \pi/50$, obtained from hydrodynamic and gravity simulations. In the gravity simulations we choose the size of the time step $\Delta t = 0.002$.
     The slopes of the reference lines (dashed and dotted lines) are computed using linear response theory.
     }
   \label{fig:grlargek}
\end{figure}

In the large $k$ regime, the results from hydrodynamic simulations do not match the results from gravity. As expected, the gradient expansion breaks down when the wave number $k \gtrsim T$.   Fig.\ \ref{fig:grlargek} shows that the relaxation time scales computed from hydrodynamics and gravity are different when $k = 20 \pi/50$.
The rest of the parameters are the same as those in the previous subsection. 

While gravity does not agree with hydrodynamics in this limit,  
the relaxation time scales computed using the full gravity simulation and its linearized counterpart do agree.
In Fig.\ \ref{fig:grlargek}, the solid black and dashed blue lines have very similar slope.  The solid black line was computed from the full gravity simulation, while the dashed blue line was computed from the low frequency limit of $\operatorname{Im} G_R^{\epsilon \epsilon}$, described in Section 4.1.2.  Equivalently, in Fig.\ \ref{fig:imG}, the star lies very nearly on the black curve.  The star was computed from the value of the slope of the solid black line in Fig.\ \ref{fig:grlargek}, while the black curve in Fig.\ \ref{fig:imG} was computed from the Green's function.  
This agreement is a non-trivial check of the gravity code.

\subsubsection*{\it 3. Dependence of relaxation time scale on $\delta$}

\begin{figure}
   \centering
 a)   \includegraphics[height=250pt]{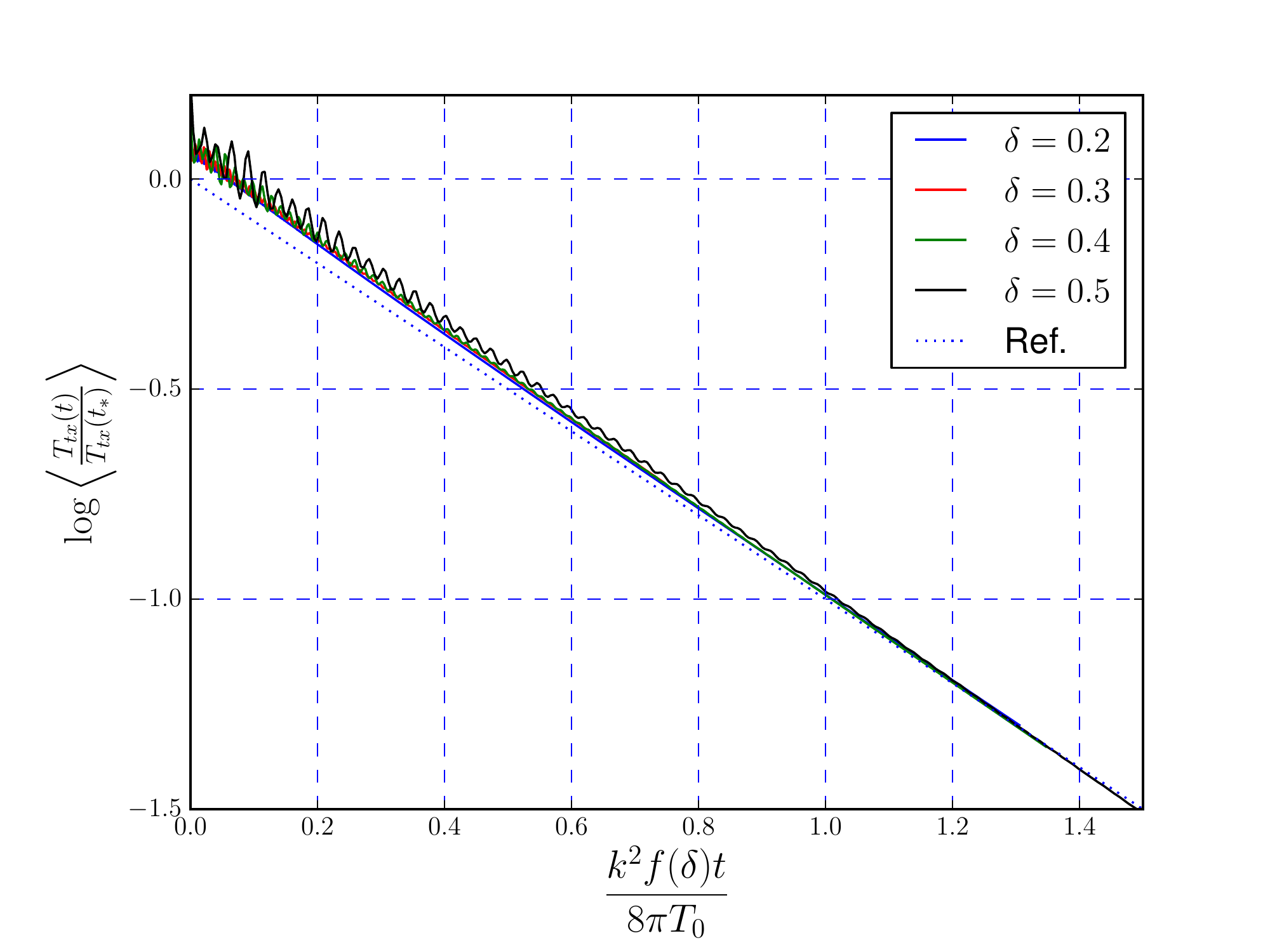}  \\ 
   b)  \includegraphics[height=250pt]{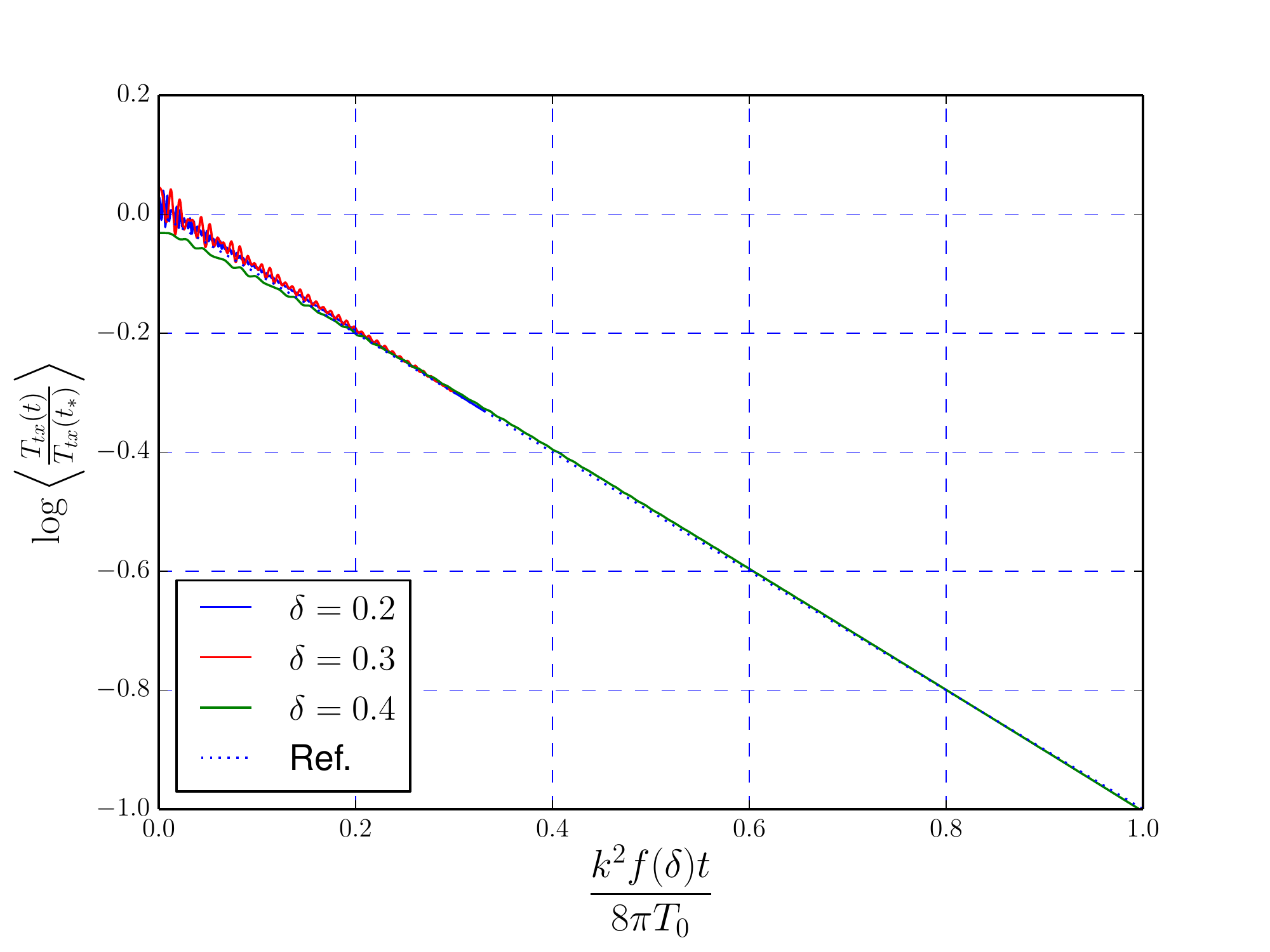} 
     \caption{A plot of $\log \langle T_{tx} \rangle$ as a function of scaled time for different values of $\delta$, obtained from a) hydrodynamic simulations and b) gravity. We have chosen $k= {\pi/50}$. The simulations were run for $25 \times 10^4$ time steps with $\Delta t =10^{-2}$. When $\delta =0.4$, the gravity simulations become less stable with $m=20$. 
     As we were only interested in the late time behavior, we were able to improve stability by choosing $m = 100$ for the $\delta = 0.4$ case.  
}
   \label{fig:HydrDelta}
\end{figure}

Here, we consider what happens in our simulations when $\delta$ leaves the linear response regime.
 In (\ref{larged}), we showed that at low wave numbers, the momentum relaxation time scale is given by
$$ {1\over \tau} = {2\eta_0 k^2 \over 3 \epsilon_0 T_0} \left( 1 - \sqrt{1-\delta^2}\right) = {\eta_0 k^2 f(\delta)\over 3 \epsilon_0 T_0} \ .$$
This expression agrees with the linear response theory computation for small $\delta$. 
To the extent to which $f(\delta) \neq \frac{\delta^2}{2}$, Fig.\ \ref{fig:HydrDelta} 
shows deviation from linear response theory for larger $\delta$. 
In particular, the results from numerical hydrodynamics and gravity show that the relaxation time scale agrees with the  low-velocity approximation result obtained analytically in (\ref{larged}), within numerical error. 

\begin{figure}[h]
   \centering
    \includegraphics[height=250pt]{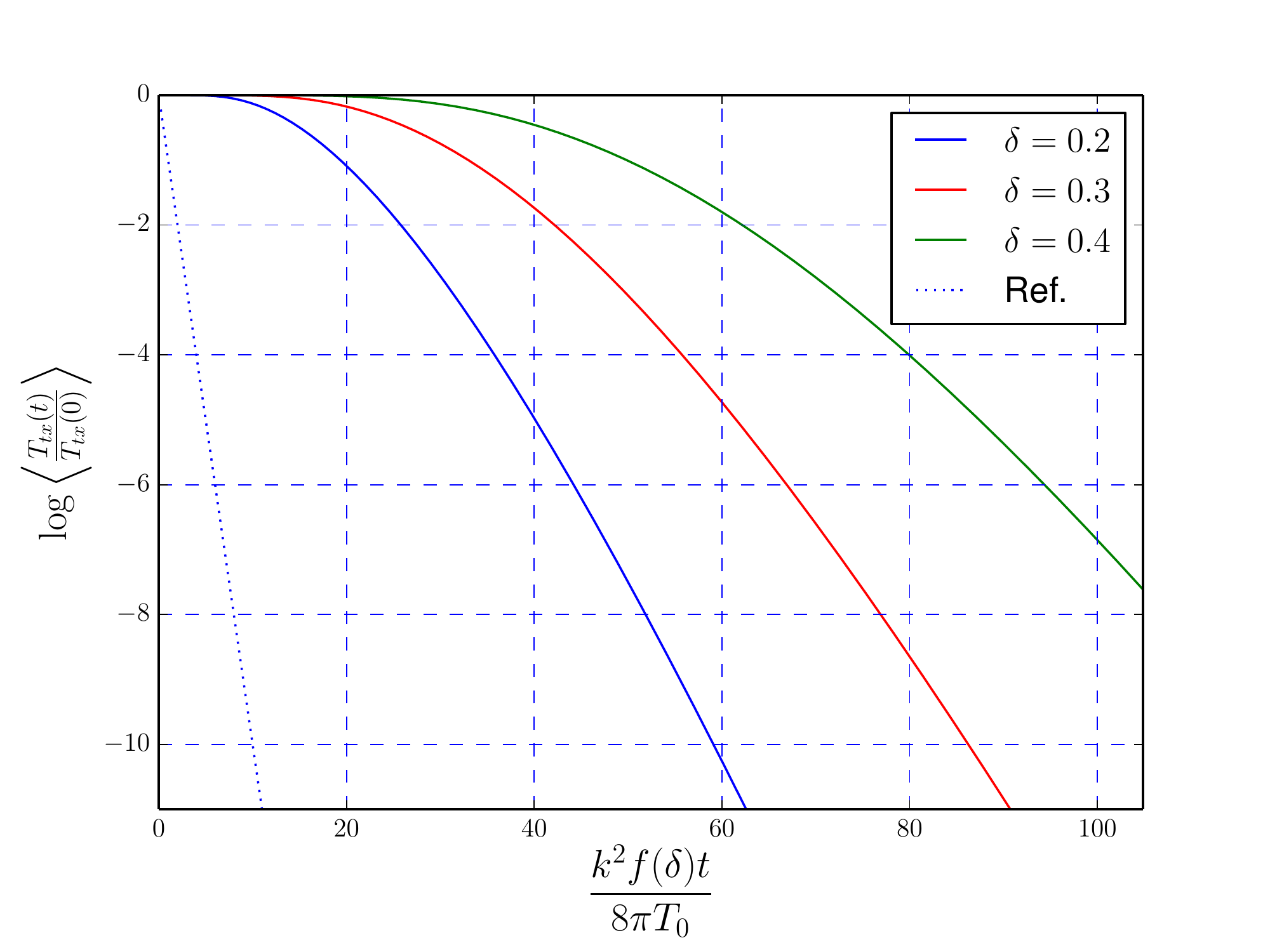} 
     \caption{ A plot of $\log \langle T_{tx} \rangle$ as a function of scaled time for different values of $\delta$, obtained from gravity simulations. We have chosen $k= {20\pi/50}$.  To improve stability, we chose
     $\Delta t =0.005$ for $\delta =0.4$, $\Delta t =0.01$ for the other two cases and $m=1000$ for all cases.
     Note $f(0.4) k^2 m /8 \pi T_0 \approx 44$.
    The reference curve is the prediction from hydrodynamics.}
   \label{fig:largedlk}
\end{figure}

The regime where $\delta$ and $k$ are both large is intractable analytically. A hydrodynamic description is not useful here as the gradient expansion (or continuum approximation) is not valid. Moreover, nonlinearities in $\delta$ are presumably important.
In Fig.\ \ref{fig:largedlk}, we show numerical results for $\delta =0.4$ and  $\delta = 0.3$ (velocity is $0.2$). 
We choose $m=1000$ to improve numerical stability. 
(There may be an issue with the development of caustics in the spacetime for smaller values of $m$ and larger values of $\delta$.)
Fig.\ \ref{fig:largedlk} shows a plot of $\log \langle T_{tx} \rangle$ versus time. This plot shows the power of our techniques to get at an otherwise inaccessible regime.  Interestingly, momentum relaxation does not appear to be exponential in time for the parameter regime explored.

Moving forward, it should be possible to obtain an analog of the low-velocity approximation (\ref{larged}) in gravity; however, the steady state solution (to which the gravity equations relax) is not known analytically when $k$ is large. One might be able to get further insight into this regime by numerically constructing the steady state regime and looking at small fluctuations about it.
Note that for large $k$, we cannot use the hydrodynamic definitions of $s$ and $T$.
Perhaps one could use the properties of the apparent horizon to define these quantities.

\section{Discussion and Outlook}
We have presented a numerical technique to study lattice-induced momentum relaxation in theories with gravity duals. We have used the characteristic formulation to solve Einstein's equations in the presence of a negative cosmological constant. We infer that the characteristic evolution scheme can be used reliably, without encountering caustics, for studying gravitational dynamics in AdS space even when metric sources are turned on. However, the numerical method becomes less stable when the strength $\delta$ of the source (lattice strength) increases.  While we have focused on momentum in this paper, one could also consider heat transport with our formalism.\footnote{%
 See ref.\ \cite{Fischetti:2012vt} for related work on this front.
}

We performed a number of cross checks of our code.
The gravity and hydrodynamic simulations agree with each other when the lattice wave number $k$ is small, i.e.\ in the hydrodynamic regime.
The relaxation time extracted from these simulations agrees with an exact analytical expression we obtained, valid for all 
$\delta$ and small $k$.  As $\delta$ is arbitrary, this agreement is not limited to the regime of linear response.  
The simulations approach a steady state equilibrium temperature distribution predicted by hydrodynamics.
Next, for large $k$ and small $\delta$, where hydrodynamics is not valid but linear response is,
the momentum relaxation time scale extracted from our gravity simulations
 agrees with the results obtained from linearized gravity. 
Given our cross checks, we can hopefully trust our numerical results in the regime of large $k$ and large $\delta$, 
where neither linear response nor hydrodynamics is valid.
Fig.\ \ref{fig:largedlk} presents some simulations of momentum relaxation in this scaling regime.

In hydrodynamics, it is straightforward to obtain analytically the solution to which the flow relaxes at late times. The answer is (\ref{Analytic}).  
On the gravity side, in principle it is straightforward to obtain this solution numerically by letting our simulations run for a long time.  
While for small $k$ this approach is feasible, we find that for large $k$, because of the exponential damping of the relaxation rate, we have to wait a prohibitively long time for the simulation to finish.  
An interesting project for the future would be to obtain the large $k$, steady-state solution directly using a different numerical approach.  This large $k$ limit is also the small temperature limit because of the relatively few scales in our problem.  The low temperature limit might conceivably be somewhat more interesting for condensed matter applications.  Of course, it would also be interesting to couple gravity to a complex scalar and gauge field.  One could then study more elaborate systems involving nonzero charge density and superfluid phases \cite{Gubser:2008px,Hartnoll:2008vx}.

 In the immediate future, we would like to use the numerical framework presented here to study more intricate flows involving shock waves, turbulent eddies, {\it etc}.  In particular, it would be interesting to understand if the gradient expansion breaks down in such flows. 

\section*{Acknowledgments}
We would like to thank P.~Chesler and L~Yaffe for discussion and for collaboration during the early stages of this project.  We thank K.~Jensen for comments on the manuscript.  
We also thank the Whiteley Center, where this project was started, for hospitality.
This work was supported in part by the National Science Foundation under Grants No.\ PHY-0844827 and PHY-1316617.  C.~H. also thanks the Sloan Foundation for partial support.

\appendix

\section{Numerical Details and Cross Checks}
\label{sec:gorydetails}

\subsection*{\it1. Overview of the numerical scheme}

In this appendix, we provide details of the numerical simulations. We begin by describing briefly pseudo spectral (collocation) methods.\footnote{%
We would like to thank P.~Chesler and L.~Yaffe for convincing us to take a spectral method approach to this problem.
} We begin by describing briefly pseudo spectral (collocation) methods. Detailed descriptions of spectral methods can be found in \cite{jpboyd,trefethen}. The basic idea of a spectral method is to expand the variables or fields in terms of basis functions $\psi_n(x)$, for example Fourier series or Chebyshev polynomials, that satisfy some orthogonality relation and are appropriate for imposing boundary conditions.
 In a pseudo spectral method, a field $u(x)$ and its gradients are evaluated at discrete collocation points $x \in \{x_0, x_1, \ldots, x_{N-1} \}$. The number of collocation points $x_j$ is the same as the number of basis functions $\psi_k$:
\begin{eqnarray}
u_N(x_j,t) \equiv \mathop{\sum}_{n=0}^{N-1} \psi_n(x_j) a_k(t),
 \quad \partial^{m}_xu_N(x,t) \equiv \mathop{\sum}_{n=0}^{N-1} {d^m \psi_n(x)\over dx^m} a_k(t),
\label{series} \end{eqnarray}
 Using orthogonality among the $\psi_n$, one can then
 express gradients of the fields in terms of linear combinations of the $u_N(x_j,t)$:
$$ \partial_x u_N(x_i,t) = \mathop{\sum}_{j=0}^{N-1}D^{(1)}_{ij}u_N(x_j,t),
\quad \partial_x^2 u_N(x_i,t) = \mathop{\sum}_{j=0}^{N-1}D^{(2)}_{ij}u_N(x_j,t),  \dots $$
 where $D_N^{(1)}$, $D_N^{(2)}, \dots$ are $N\times N$ derivative matrices.
 
 In this paper, our PDEs are all schematically of the form
$\partial_t u = {\cal L}(u,\partial_x u, $ $\partial_x^2 u, \dots)$.  To a first approximation, we solve such an equation by first replacing fields and spatial derivative operators with their discrete versions $u_N$, $D_N^{(1)}$, $D_N^{(2)}$, etc., and then
integrating in time using a Runge-Kutta or Adams-Bashforth scheme.
The relativistic hydrodynamic equations take precisely this form, treating $u$ as a five dimensional vector with components $(T, u^x, u^y, B, \Pi^{xy})$.
We work in a box with periodic boundary conditions, and thus we use Fourier basis functions $\psi_n(x) = e^{inx}$ to compute the derivatives appearing on the RHS of the hydrodynamic equations.   We use the third order Adams-Bashforth technique to integrate in time.
 The starting values at $t = 0, \Delta t, 2\Delta t$ are computed using Runge-Kutta integration. The error in this method of integration is ${\cal O}\left(\Delta t^3\right)$.  As the equations already contain viscous terms, we do not need to add artificial viscosity or use filtering to stabilize the code.

We will now proceed to describe the details of the gravity simulations. We use a coordinate system such that the apparent horizon is at $z=1$ and the boundary is at $z=0$. We discretize the holographic direction using a Chebyshev grid and the boundary spatial directions using a Fourier grid.
As described in the text, the characteristic method allows us to write Einstein's equation in a form that has a nested structure.
The discretized Einstein's equations take the following form; 
\begin{eqnarray}
 \left(D_z + {4\over z} \mathbb{I} \right) \beta_s &=& S_{\beta} \left( z, \alpha_s,  \theta_s,\chi_s\right)
 \label{beq}
\\
\left(D_z  \right) {\pi^A_s} &=& S_{\pi^A} \left( z, \alpha_s, \theta_s,\chi_s, \beta_s \right)
 \label{PiUeq}
 \\
 \left(D_z + {2\over z} \mathbb{I} \right) {U^A}_s &=& S_{U^A} \left( z, \alpha_s,  \theta_s,\chi_s, \beta_s, \pi^A_s \right) \label{Ueq}
 \\
\left(D_z  \right) {d_t \chi}_s &=& S_{{d_t\chi}} \left( z,  \alpha_s,  \theta_s,\chi_s, \beta_s, U^A_s \right)
 \label{dtchieq}
 \end{eqnarray}
 \begin{eqnarray}
\left(D_z + {1\over z} \mathbb{I} \right) {d_t \alpha}_s + C_{\alpha \alpha} {d_t \alpha}_s + C_{\alpha \theta} {d_t \theta}_s &=& S_ {{d_t\alpha}}\left(\dots,d_t\chi_s\right)
 \label{dtalphaeq}
 \\
 \left(D_z + {1\over z} \mathbb{I} \right) {d_t \theta}_s + C_{\theta \alpha} {d_t \alpha}_s + C_{\theta \theta} {d_t \theta}_s &=& S_ {{d_t\theta}}\left(\dots,d_t\chi_s\right)
  \label{dttheq}
 \end{eqnarray}
\begin{eqnarray}C^H_{xx} D^{(2)}_x V_H + C^H_x D^{(1)}_x V_H + C^H_0 V_H = S_{{V_H}}\left(\alpha_H, \theta_H, \beta_H,U^A_H,\chi_H,d_t\chi_H, d_t \alpha_H, d_t \theta_H \right) \label{appeq}  \end{eqnarray}
\begin{eqnarray}
\partial_t \alpha_s &=& {1\over z} (d_t \alpha)_s + {1\over 2} {\left(z \alpha'_s + 2 \alpha_s\right) }z^3 V \label{aldoteq} \\
\partial_t \theta_s &=& {1\over z} (d_t \theta)_s + {1\over 2} {\left(z \theta'_s + 2 \theta_s\right) } z^3 V \label{thdoteq} \\
\partial_t \chi &=& S_\chi  \left(V_H, U^A_H, \chi_H \right)  \label{dtchieq2} \\
\partial_t U^{A}_3 &=& S_{U^{A}_3} \left( \alpha_3, \theta_3, \chi_3, V_3, U^{A}_3,  \beta_0\right) \label{ubdyeq} \\
\partial_t V_3 &=& S_{V_3} \left( \alpha_3, \theta_3, \chi_3, V_3, U^{A}_3,  \beta_0\right) \label{W3eq}
\end{eqnarray}
Note that time derivatives of $X_i$ do not appear in the source terms $S(X_1,\dots,X_n)$ but spatial gradients generically will.

As described in section \ref{sec:numericalGR}, 
eqs.\ (\ref{beq})-(\ref{appeq}) are hypersurface equations {\cite{Winicour}}, i.e.\ 
we solve these equations at every time slice and propagate this information to the next time 
using the bulk evolution equations (\ref{aldoteq}), (\ref{thdoteq}) and the boundary evolution 
equations (\ref{dtchieq2}) and (\ref{ubdyeq}). The time constraint equations equations 
(\ref{beq})-(\ref{appeq}) must be satisfied at $t=0$ in order to have well defined Cauchy data. 
The Cauchy data consist of initial conditions for $\alpha(x,z,t=0)$, $\theta(x,z,t=0)$, 
 $U^A_3(x,t=0)$ and $\chi_3(x,t=0)$. 
 Numerically, the boundary conditions are applied by replacing the first row of the matrices appearing on the LHS of the time constraint equations (\ref{beq})-(\ref{dttheq}) and the first entry of the source terms.
 Using Fourier series, periodic boundary conditions for the elliptic eq.\ (\ref{appeq}) are enforced automatically.  
 Moreover, generically
 (\ref{appeq}) will have no zero modes.
To integrate the time evolution equations, we have used both third order Adams-Bashforth and fourth order Runge-Kutta.
As a consequence of the Bianchi identities (see section 2), the last equation (\ref{W3eq}) should be satisfied if all other equations are satisfied. We use this Bianchi constraint for monitoring the error, as we discuss in greater detail later in this section.

Note that the operators appearing on the LHS of equations (\ref{beq})-(\ref{dtchieq})
are independent of $x$ and $t$. These equations can be solved by multiplying the source terms by the inverse of these operators. It is sufficient to compute the inverse of these operators only once, leading to an improvement in the speed of the code. As mentioned earlier, we choose Neumann boundary condition for $\beta_s$ and $U^A_s$ to minimize the condition numbers of the matrices appearing on the LHS of (\ref{beq}) and (\ref{Ueq}). The round-off error arising from the inversion of operators with lower condition number is less.

There are several unpleasant details involved in implementing this scheme that remain to be discussed.

\subsection*{\it 2. Factors affecting stability and accuracy of the code}
There are many factors that affect the reliability of numerical methods. We discuss some of the issues that we did and did not encounter and the strategies we employed to make the code work.
\begin{itemize}
\item {\it Round-off}

We encountered round-off error in trying to evaluate the source terms in the PDEs for our gravity code.
These source terms are large rational and polynomial expressions.
One technique that helped reduce error was to place the polynomials in
Horner form.\footnote{%
In Mathematica, the relevant command is {\tt HornerForm}.
 }
 Another strategy was judicious use of $\tt{expm1}$ and $\tt{log1p}$ functions that avoid round-off error in evaluating $e^x - 1$ and $\log(1+x)$ respectively when $x$ is close to zero.  A typical situation was a need to evaluate accurately an expression like $(e^z - 1) / z$ close to the conformal boundary $z=0$.

\item{\it Truncation error}

Truncation error, or the error associated with the discretization, is typically less of an issue for spectral methods then for finite volume methods.
We found in our simulations that the truncation error in the $z$-direction became comparable to machine epsilon
with fewer than 20 collocation points.  In the $x$-direction, we needed in contrast on the order of 100 points to achieve accuracy at the part per million level.

There is also a truncation error arising from time integration.
We found the third order Adams-Bashforth technique sufficient for our computation, resulting in an ${\cal O}\left(\Delta t^3 \right)$ error in the constraints.  The error can be ameliorated by using a fourth order Runge-Kutta or Adams-Bashforth method.

\item {\it Aliasing error}

While viscous terms eliminate the need for filtering the hydrodynamic code, aliasing is a noticeable problem for the gravity code.
The standard solution to this problem, which we have implemented, is a low pass filter.
The filtering leads to significant improvement in the behavior of constraints and stability of the code.
We implement filtering in real space using matrix multiplication.

For the Fourier grid, we employ a 2/3-rule.
The fields take the schematic form
$$ u(x) = \mathop{\sum}_j h(x,x_j) u(x_j)$$
where, $$h(x,x_j) = {1\over N} \sum_{n=-N/2}^{N/2} {1\over c_n}e^{i n (x-x_j)}, \quad {\text{with }}c_n = 1 + \delta_{|n|,N/2}$$
In order to get rid of the higher modes we use a low-pass filter by computing $u(x)$ on a coarser grid with 1/3 of the modes and 1/3 of the collocation points eliminated:
$$ \tilde{u}(\tilde{x}_i) = \sum_j{1\over N} \mathop{\sum}_{n=-N/2}^{N/2} {1\over c_n}\sigma \left({2|n|\over N}\right) e^{i n (\tilde x_i-x_j)} u(x_j) =\sum_j {\cal F}^N_{ij} u (x_j)$$
where $\sigma(|n|/N) = 0 \text{ for }|n| > N/3$ and one otherwise.   ${\cal F}^{N}$ is the ``filtering-matrix". We then interpolate back to a finer grid using the interpolating matrix ${\cal I}^N$ which is obtained by setting $\sigma(x) = 1$. The de-aliasing matrix is product of ${\cal I}^N$ and ${\cal F}^N$.

Our radial filtering for the Chebyshev grid just involves interpolating to a coarser grid and then back to the original grid, that is, we choose the filter function $\sigma$ to be 1 for Chebyshev methods.

\item {\it Formation of caustics}

As discussed in section 2, the utility of the null-characteristic formulation will break down in the presence of caustics,
and our numerical scheme will fail.  Fortunately we did not encounter caustics for a wide range of parameters.  The gravity code is less stable for very large $\delta \gtrsim 0.5$ suggesting a possible formation of caustics in this regime.

\end{itemize}

\subsection*{\it 3. Some checks on the numerical method}

\begin{figure}
   \centering
    \includegraphics[height=270pt]{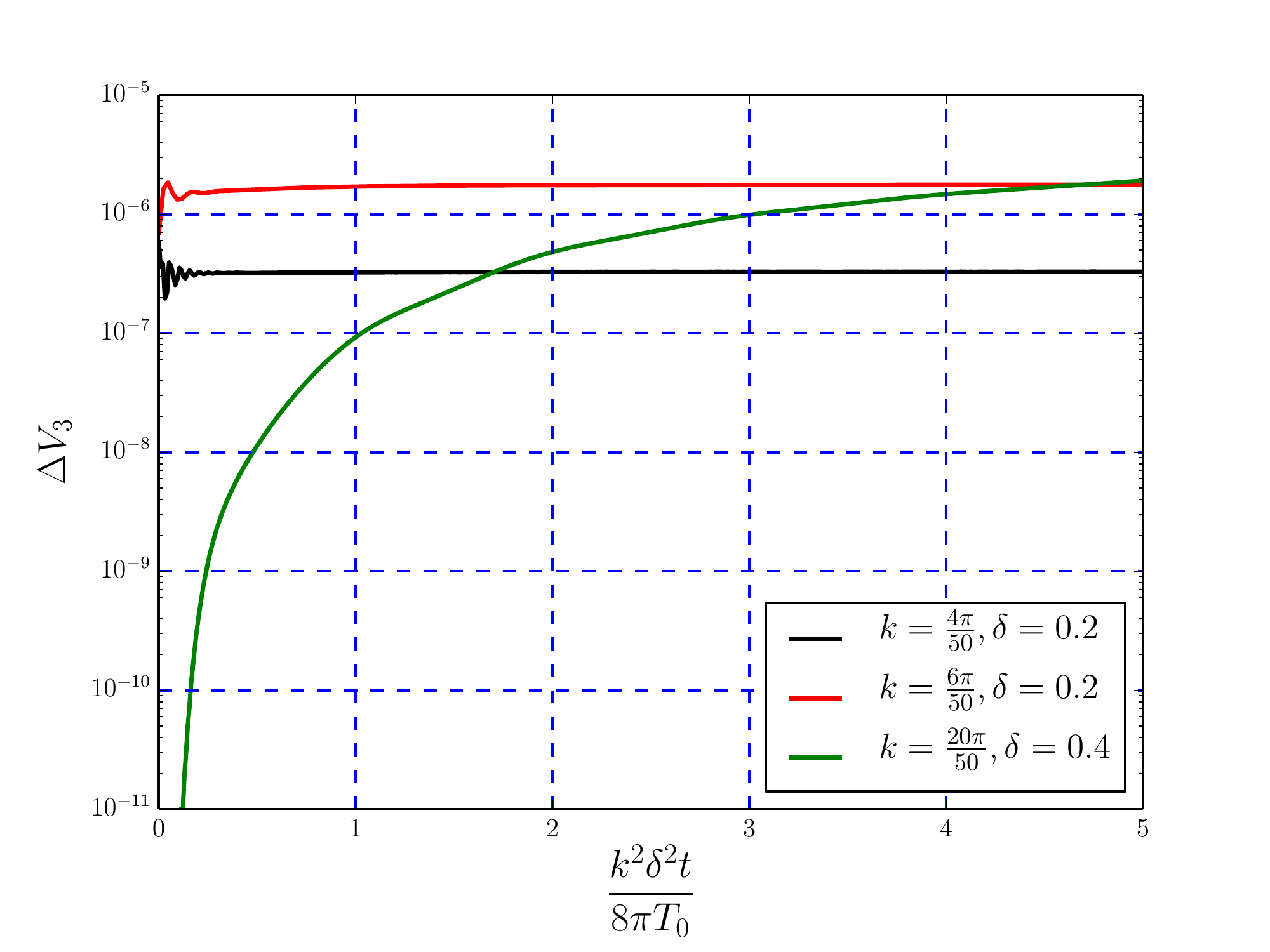} 
     \caption{ A plot of $\Delta V_3$ as a function of scaled time for different values of  $k$ and $\delta$.}
   \label{fig:constraint}
\end{figure}

 It is important that the constraints arising from Bianchi identities (discussed earlier) remain close to zero. We check the Bianchi constraint by computing $V_3$ in two different ways: (i) using equations (\ref{dtchieq}),
(\ref{appeq}) and (\ref{dtchieq2}),\footnote{Recall that $\chi = {1\over 4} \log
 \left(1+ 2 z^3 \chi_3\right)$. } and (ii) using the boundary equation (\ref{W3eq}).
 These two methods of evaluating $V_3$ should give the same result up to
numerical error.
In all our numerical simulations the constraints remained around $10^{-6}$ or $10^{-7}$ in the presence of radial filtering. When radial filtering was turned off the constraints became as large as $10^{-4}$ but eventually decreased to $10^{-6}$. Fig.\ \ref{fig:constraint} shows that the constraints remain within $10^{-6}$ for some representative cases.

\begin{figure}     
 \centering
          \includegraphics[height=270pt]{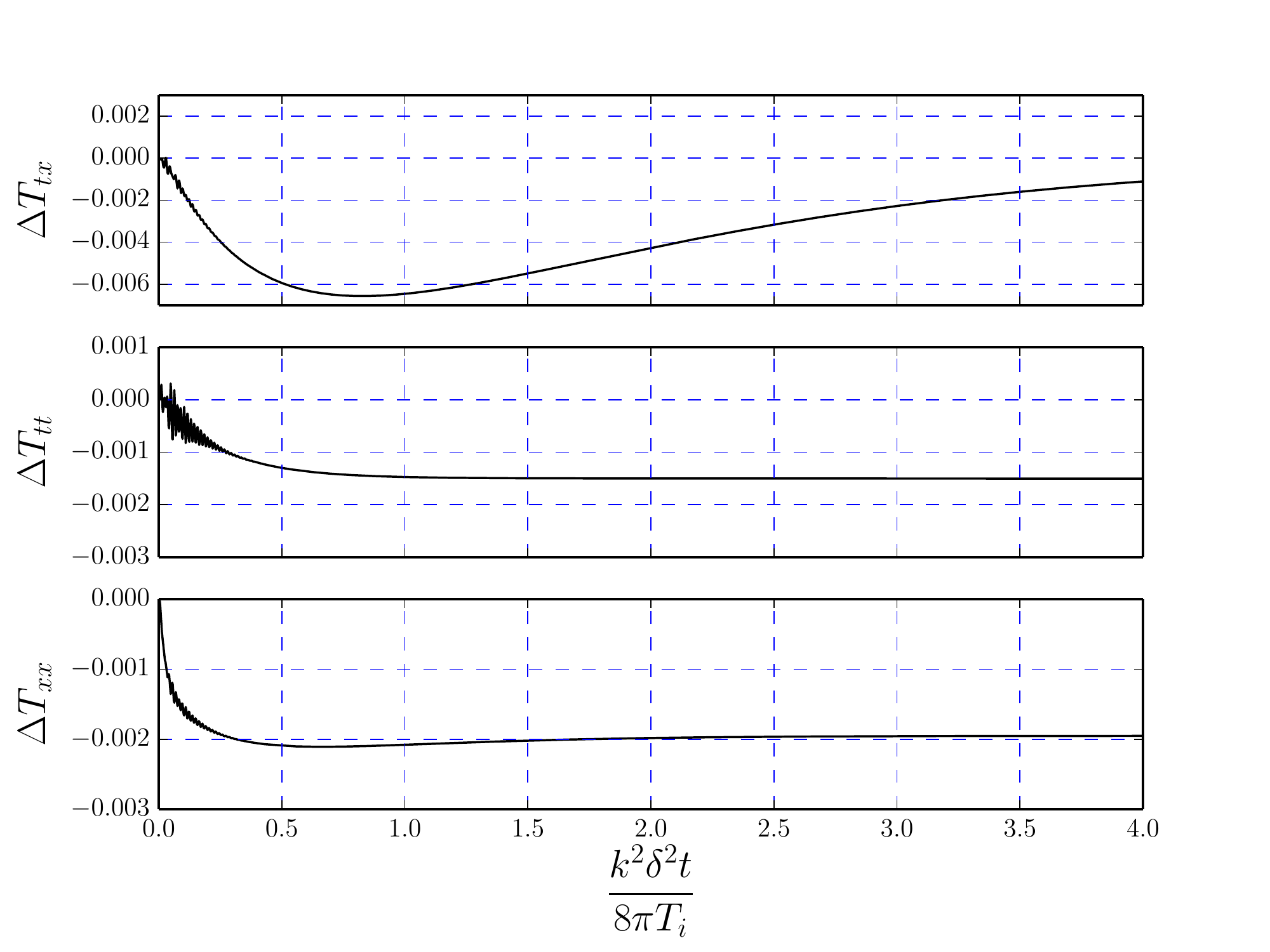} 
     \caption{Plots of  $\Delta T_{\mu\nu}$ obtained from hydrodynamic and gravity simulations with $k = 4 \pi / 50$ and 
     $\delta = 0.2$.}
   \label{fig:diffStress}
\end{figure}

Another non-trivial check of our method is to compare the hydrodynamic stress-tensor with the stress-tensor obtained using gravity for low values of $k$ (hydrodynamic regime). Fig.\ \ref{fig:diffStress} show that the difference in $T_{tx}$, $T_{tt}$ and $T_{xx}$ when $k = 4\pi/50$ is around $10^{-3} \sim {\cal O}(k^5)$.

\subsection*{\it 4. Computational Platform}

Two separate software packages were developed to run the simulations described in this paper.  The first
was developed using Matlab \cite{matlab} and used the code described in ref.\ \cite{Chesler:2007an} as a starting point.\footnote{%
We thank Paul Chesler for making his code available to us.
}
The second was developed from scratch using Python.  
The simulations were run on standard desktop and laptop computers.  The longest gravity simulation (1 million time steps) took about 6 hours to complete.
The Python code ultimately ran about ten times faster than the Matlab code although at some point (largely for fiscal reasons) we stopped trying to optimize the Matlab code.
%

\section{Small $\delta$ momentum relaxation from hydrodynamics}
\label{sec:twolimits}

In this section, we present a method using hydrodynamics to compute the relaxation time scale when the metric perturbation is small but the velocity is arbitrary. This method does not rely on the memory function formalism, but the results agree with (\ref{linresp}) when the velocity is small. 
We look for perturbative solutions of the relativistic hydrodynamic equations with the structure
\begin{eqnarray}
g_{tt} &=& -(1 +  2 g_1 \cos(k x) \delta)  \ , \\
T &=& T_0(t) +  (T_1 e^{i k x } + T_1^* e^{-ikx}) \delta +  (T_2 e^{2i k x } + T_2^* e^{-2ikx}) \delta^2 + O(\delta^3) \ , \\
u^1 &=&  v(t) + (v_1 e^{i k x} + v_1^* e^{-i k x})  \delta + (v_2 e^{2i k x} + v_2^* e^{-2i k x})  \delta^2+ O(\delta^3) \ .
\end{eqnarray}
We will allow $T_0(t)$ and $v(t)$ to be weakly time dependent, i.e.\ $\dot T_0 \sim \dot v \sim \delta^2$. {Let us define $\dot{\tilde T}_0 = \delta^{-2} \dot T_0$ and $\dot {\tilde v} = \delta^{-2} \dot v$.}
We first solve for $T_1$ and $v_1$:
\begin{eqnarray}
T_1 &=& - \frac{c_s^2 g_1 s T^2}{2 (c_s^2 s T(1 + v^2) + i v( i s T v + k (1+v^2) (\eta+\xi)))} \ , \\
v_1 &=& - \frac{ g_1 v (1+v^2) ( (c_s^2-1) s T + i k v (\eta+\xi))}{2(c_s^2 s T(1+v^2) + i v (i s T v + k (1+v^2)(\eta + \xi)))} \ .
\end{eqnarray}
We have made use of the 
thermodynamic relation $\epsilon + p = s T$ where $s$ is the entropy density and introduced the speed of sound $c_s^2 = \partial p / \partial \epsilon$.

We then calculate how well the stress tensor conservation condition is met at second order in our expansion.
The $x$-dependent pieces will be satisfied by adjusting $T_2$ and $v_2$ accordingly.  The $x$-independent piece
will allow us to solve for the time dependence of $T_0(t)$ and $v(t)$.  
Let us first consider the energy conservation condition.
The $x$-independent piece at second order (equivalently the spatially averaged piece) is
\begin{eqnarray}
\label{aveE}
\langle \nabla_\mu T^{\mu 0} \rangle &=& -\left(2 s T v \dot {\tilde{v}} + \frac{s}{c_s^2} (1 + (1+c_s^2) v^2) \dot {\tilde{T}} \right) \delta^2  + \ldots \\ 
&=&- \frac{d}{dt} \left[ s T v^2 + \epsilon \right] \ . 
\end{eqnarray}
This equation can be trivially integrated and expresses energy conservation
\begin{eqnarray}s T v^2 + \epsilon = \rm{constant} \ .
\label{Econs}
\end{eqnarray}
For momentum conservation, we find (using the relation for $\dot T$ from energy conservation)
\begin{eqnarray}
\label{aveP}
\langle \nabla_\mu T^{\mu 1} \rangle &=& 
\frac{d}{dt} s T v \sqrt{1+v^2}  + \\
&&
+ \frac{g_1^2 k^2 s^2 T^2 (\eta+\xi) v}{2 | c_s^2 s T(1+v^2) + i v(i s T v + k(1+v^2) (\eta + \xi))|^2}  \delta^2 + \ldots
\nonumber \ , \\
&=&
s T \Biggl[
\left(\frac{2(1+v^2)}{1 + (1+c_s^2) v^2} - 1\right) \frac{\dot {\tilde{v}}}{\sqrt{1+v^2}} + \\
&&
+ \frac{g_1^2 k^2 s T (\eta+\xi) v}{2 | c_s^2 s T(1+v^2) + i v(i s T v + k(1+v^2) (\eta + \xi))|^2} \Biggr] \delta^2 + \ldots
\nonumber \ .
\end{eqnarray}
At leading order in velocity, momentum conservation reduces to
\begin{eqnarray}
\label{avePsimp}
\langle \nabla_\mu T^{\mu 1} \rangle = s T \dot v +   \frac{(\eta + \xi)v}{2} \left(\frac{g_1 k \delta }{c_s^2} \right)^2  + O(v^3, \delta^3) \ .
\end{eqnarray}
From equations (\ref{Econs}) and (\ref{aveP}), it is clear that $v$ decreases slowly over time while $T$, by energy conservation, must increase.

In the conformal case, where $c_s^2 = 1/2$ and $\xi = 0$, we calculate the form of the decrease in velocity and increase in temperature. 
We parametrize the energy density and viscosity by $\epsilon = \epsilon_0 T^3$ and $\eta = \eta_0 T^2$.  
Thus $s T = 3 \epsilon_0 T^3 /2$.  Assuming first that $T$ is roughly constant, we find that
 \[
v \approx v_0 \exp \left( 
\frac{- 4\eta_0 g_1^2 k^2 \delta^2}{3 \epsilon_0 T} t 
 \right) \ .
 \]
Then from energy conservation it follows that 
 \begin{eqnarray}
 T^3 \approx T_0^3 \frac{v_0^2 + 2/3}{v^2 +2/3} \ ,
 \label{Tvariation}
 \end{eqnarray}
 where $T_0$ is the initial temperature.

\end{document}